\documentclass[twocolumn]{aastex62} 

\newcommand{\teff}{$T_{\rm{eff}}$}
\newcommand{\chisq}{$\chi^2$\ }
\newcommand{\kms}{km~s$^{-1}$}
\usepackage{xcolor}	
\usepackage{graphicx}	

\begin{document}

\title{Visual Orbits of Spectroscopic Binaries with the CHARA Array. III. HD~8374 and HD~24546}

\author[0000-0002-9903-9911]{Kathryn V. Lester}
\affil{Center for High Angular Resolution Astronomy and Department of Physics \& Astronomy, \\ Georgia State University, Atlanta, GA 30302 USA}

\author[0000-0002-9413-3896]{Francis C. Fekel}
\affil{Center of Excellence in Information Systems, Tennessee State University, Nashville, TN 37209 USA}

\author{Matthew Muterspaugh}
\affil{Division of Science, Technology, \& Mathematics, Columbia State Community College, Columbia, TN 38401 USA}

\author[0000-0001-8537-3583]{Douglas R. Gies}
\affil{Center for High Angular Resolution Astronomy and Department of Physics \& Astronomy, \\ Georgia State University, Atlanta, GA 30302 USA}

\author[0000-0001-5415-9189]{Gail H. Schaefer}  
\affil{The CHARA Array of Georgia State University, Mount Wilson Observatory, Mount Wilson, CA 91023 USA}

\author[0000-0001-9939-2830]{Christopher D. Farrington}
\affil{The CHARA Array of Georgia State University, Mount Wilson Observatory, Mount Wilson, CA 91023 USA}

\author[0000-0002-0951-2171]{Zhao Guo}   		
\affil{Department of Astronomy \& Astrophysics, Pennsylvania State University, University Park, PA 16802 USA}

\author[0000-0001-7233-7508]{Rachel A. Matson}  
\affil{NASA Ames Research Center, Moffett Field, CA 94035 USA}
\affil{U.S. Naval Observatory, Washington, DC 20392 USA}

\author[0000-0002-3380-3307]{John D. Monnier}  
\affil{Department of Astronomy, University of Michigan, Ann Arbor, MI 48109 USA}

\author[0000-0002-0114-7915]{Theo ten Brummelaar}  
\affil{The CHARA Array of Georgia State University, Mount Wilson Observatory, Mount Wilson, CA 91023 USA}

\author{Judit Sturmann}  
\affil{The CHARA Array of Georgia State University, Mount Wilson Observatory, Mount Wilson, CA 91023 USA}

\author{Samuel A. Weiss} 
\affil{Department of Physics, Southern Connecticut State University,  New Haven, CT 06515 USA}

\correspondingauthor{Kathryn Lester}
\email{lester@astro.gsu.edu}

\begin{abstract}

We present the visual orbits of two long period spectroscopic binary stars, HD~8374 and HD~24546, using interferometric observations acquired with the CHARA Array and the Palomar Testbed Interferometer. We also obtained new radial velocities from echelle spectra using the APO 3.5~m and Fairborn 2.0~m telescopes. By combining the visual and spectroscopic observations, we solve for the full, three-dimensional orbits and determine the stellar masses and distances to within 3\% uncertainty. We then estimate the effective temperature and radius of each component star through Doppler tomography and spectral energy distribution analyses, in order to compare the observed stellar parameters to the predictions of stellar evolution models. 
For HD~8374, we find 
masses of 
$M_1 = 1.636\pm0.050 M_{\odot}$ and 
$M_2 = 1.587\pm0.049 M_\odot$, 
radii of 
$R_1 = 1.84\pm0.05 R_\odot$ and 
$R_2 = 1.66\pm0.12 R_\odot$, 
temperatures of 
$T_{\rm eff \ 1} = 7280\pm110$ K and 
$T_{\rm eff \ 2} = 7280\pm120$ K, 
and an estimated age of 1.0 Gyr.
For HD~24546, we find
masses of 
$M_1 = 1.434\pm0.014 M_{\odot}$ and 
$M_2 = 1.409\pm0.014 M_\odot$, 
radii of 
$R_1 = 1.67\pm0.06 R_\odot$ and 
$R_2 = 1.60\pm0.10 R_\odot$, 
temperatures of 
$T_{\rm eff \ 1} = 6790\pm120$ K and 
$T_{\rm eff \ 2} = 6770\pm90$ K, 
and an estimated age of 1.4 Gyr. HD~24546 is therefore too old to be a member of the Hyades cluster, despite its physical proximity to the group.
\end{abstract}


\section{Introduction}
We are continuing our series of papers measuring the visual orbits of spectroscopic binary stars with interferometry \citep{lester19a, lester19b}, in order to determine the fundamental stellar parameters of the components and test the predictions of stellar evolution models.  In this paper, we present the results for two long period binaries, HD~8374 and HD~24546. HD~8374\footnote{47 And, HR 395;  $V=5.6$ mag; $\alpha=01:23:40.6$,  $\delta=+37:42:53.8$  (J2000)} contains a pair of late A-type stars with an orbital period of 35 days. The first spectroscopic orbit was completed by \citet{fletcher67}, then recently updated using high resolution data by \citet{fekel11}. HD~8374 is also classified as a metallic line (Am) star due to its weak \ion{Ca}{2} H \& K lines \citep{abt95}. 

HD~24546\footnote{43 Per A, BD +50 860, HR 1210; $V=5.3$ mag; $\alpha=03:56:36.6$,  $\delta= +50:41:43.4$ (J2000)} contains a pair of F5 stars with an orbital period of 30 days. Spectroscopic orbits were determined by \citet{wallerstein73} and \citet{abt76}. HD~24546 also has a possible third companion (43 Per B\footnote{43 Per B, BD +50 861;  $V=10.5$ mag; $\alpha=03:56:40.7, \ \delta=+50:42:47.5$ (J2000)}) at a separation of 75\arcsec\ \citep{abt76, tokovinin97} with a proper motion and parallax similar to HD~24546 \citep{lepine07, montes18}. Even if 43~Per~B is physically associated with HD~24546, it is outside the field-of-view of our telescopes and would not cause perturbations in the orbit of HD~24546 due to the estimated orbital period of 95,000 years \citep{tokovinin97}. Some investigators have reported that HD~24546 is a member of the Hyades cluster \citep{eggen71, montes18}, while others found that it is not \citep{perryman98}, so measuring accurate masses and age for this system will verify or refute cluster membership.

In Section~\ref{section:spec}, we describe our spectroscopic observations and radial velocities. In Section~\ref{section:inter}, we describe our interferometric observations, binary positions, and combined orbital solution. In Section~\ref{section:param}, we determine the fundamental stellar parameters of each component and compare the results to stellar evolution models. We further discuss our results in Section~\ref{section:discussion}.

\section{Spectroscopy}\label{section:spec}

\subsection{APO Observations}
We observed HD~8374 and HD~24546 with the ARC echelle spectrograph \citep[ARCES;][]{arces} on the APO 3.5~m telescope from 2015--2020. Our observations are listed in Table~\ref{rvtable8374} for HD~8374 and Table~\ref{rvtable24546} for HD~24546. ARCES covers 3500-10500 \AA\ over 107 orders at an average resolving power of $R=30,000$. We reduced our data using standard echelle procedures in IRAF, then removed the blaze function using the procedure in Appendix~A of \citet{kolbas15}. 
Radial velocities ($V_r$) for the APO data were measured with the multi-order TODCOR method  \citep{todcor1, todcor2} as described in \citet{lester19a}. Briefly, TODCOR calculates the cross correlation function (CCF) for a grid of primary and secondary radial velocities. BLUERED model spectra \citep{bluered} were created as template spectra based on atmospheric parameters from \citet{fekel11} for HD~8374 and from \citet{wallerstein73} for HD~24546. The CCFs for each echelle order were then added together to find the maximum CCF and corresponding best-fit radial velocities.  TODCOR also estimates the monochromatic flux ratio near H$\alpha$ to be $f_2 / f_1 = 0.91 \pm 0.12$ for HD~8374 and $f_2 / f_1 = 0.98 \pm 0.10$ for HD~24546.

\subsection{Fairborn Observations}
We also acquired spectroscopic observations of HD~8374 and HD~24546 at Fairborn Observatory in southeast Arizona with the Tennessee State University 2.0 m Automatic Spectroscopic Telescope (AST) and a fiber-fed echelle spectrograph \citep{eaton04}. The new AST observations of HD~8374 from 2011-- 2019 are listed in Table~\ref{rvtable8374} and are a continuation of those published by \citet{fekel11}. The observations of HD~24546 acquired between 2003-- 2019 are listed in Table~\ref{rvtable24546}. All observations through the spring of 2011 were acquired with a 2048 $\times$ 4096 SITe ST-002A CCD. Those spectra have 21 orders, cover a wavelength region of 4920--7100~\AA, and have a resolving power of 35,000 at 6000~\AA. During the summer of 2011 we replaced the SITe CCD with a Fairchild 486 CCD that has a 4096 $\times$ 4096 pixel array enabling coverage of a wavelength range of 3800--8600~\AA\ over 48 orders \citep{fekel13}. We used a 200~$\mu$m fiber that produced a resolving power of 25,000 at 6000~\AA. \citet{eaton07} explained the reduction and wavelength calibration of the raw AST spectra.

\citet{fekel09} provided a general description of the typical velocity reduction. For the two stars in this work, we used a solar-type star line list consisting of 168 lines in the wavelength region 4920--7100~\AA. Each line was fitted with a rotational broadening function \citep{sandberg11}, and when the lines of the two components were blended we obtained a simultaneous fit. The stellar velocity was determined as the average of the line fits. A value of 0.3 km~s$^{-1}$ was added to the SITe CCD velocities and 0.6 km~s$^{-1}$ to the Fairchild CCD velocities to make the resulting velocities from the two CCDs consistent with the velocity zero point of \citet{scarfe10}. 

\subsection{Preliminary Spectroscopic Orbit} \label{sborbit}
To account for differences in the zero-point offsets of the APO and Fairborn spectrographs, we first fit separate orbital solutions to each data set using the RVFIT program\footnote{\href{http://www.cefca.es/people/~riglesias/rvfit.html}{http://www.cefca.es/people/$\sim$riglesias/rvfit.html}} \citep{rvfit} to solve for the spectroscopic orbital parameters: the orbital period ($P$), epoch of periastron ($T$), eccentricity ($e$), longitude of periastron of the primary star ($\omega_1$), systemic velocity ($\gamma$), and the velocity semi-amplitudes ($K_1$, $K_2$). We found offsets of $-0.14$ \kms\  and $-0.20$ \kms\ for the ARCES data of HD~8374 and HD~24546, respectively, in order to match the systemic velocities to those of the Fairborn data. We also fit an orbit to the previously published velocities of HD~8374 from \citet{fekel11} using preliminary uncertainties equal to $1/\sqrt{weight}$. We then used the \chisq\ values from the individual APO, Fairborn, and \citet{fekel11} solutions to rescale the uncertainties such that the reduced $\chi^2$ of each data set equals one. The adjusted APO velocities and the rescaled uncertainties for the APO and new Fairborn data are listed in Tables \ref{rvtable8374} and \ref{rvtable24546}, along with the residuals from the combined (VB+SB2) orbital solutions found in Section~\ref{vbsbfit}.

\begin{deluxetable*}{lccrrrrrrc}
\tablewidth{0pt}
\tabletypesize{\footnotesize}
\tablecaption{  Radial Velocity Measurements for HD~8374\label{rvtable8374}    }
\tablehead{ \colhead{UT Date} & \colhead{HJD-2,400,000} & \colhead{Orbital} & \colhead{$V_{r1}$} 
& \colhead{$\sigma_1$} & \colhead{$O-C$}  & \colhead{$V_{r2}$} & \colhead{$\sigma_2$} & \colhead{$O-C$} & \colhead{Source}
\\  
\colhead{} & \colhead{} & \colhead{Phase} & \colhead{(km~s$^{-1}$)} & \colhead{(km~s$^{-1}$)} 
& \colhead{(km~s$^{-1}$)} & \colhead{(km~s$^{-1}$)} & \colhead{(km~s$^{-1}$)} & \colhead{(km~s$^{-1}$)} & \colhead{}   }
\startdata 
 2011 Mar 03  & 55623.6328  &  0.61  &  $  -1.60$  &    0.32  &  $ -0.41$  &  $  29.80$  &    0.29  &  $ -0.15$  &   Fairborn \\ 
2011 May 20  & 55701.9727  &  0.83  &  $  -3.70$  &    0.32  &  $  0.03$  &  $  32.80$  &    0.29  &  $  0.24$  &   Fairborn \\ 
2011 Oct 11  & 55845.6211  &  0.89  &  $   0.10$  &    0.32  &  $ -0.22$  &  $  28.30$  &    0.29  &  $ -0.08$  &   Fairborn \\ 
2011 Dec 27  & 55922.5820  &  0.06  &  $  53.50$  &    0.32  &  $  0.16$  &  $ -26.50$  &    0.29  &  $ -0.24$  &   Fairborn \\ 
2012 Jun 21  & 56099.9141  &  0.08  &  $  48.10$  &    0.32  &  $  0.15$  &  $ -20.60$  &    0.29  &  $  0.11$  &   Fairborn \\ 
 2012 Sep 01  & 56171.9648  &  0.12  &  $  37.20$  &    0.32  &  $  0.20$  &  $  -9.70$  &    0.29  &  $ -0.28$  &   Fairborn \\ 
 2012 Oct 02  & 56202.6953  &  0.98  &  $  55.10$  &    0.32  &  $  0.11$  &  $ -27.90$  &    0.29  &  $  0.06$  &   Fairborn \\ 
2012 Oct 30  & 56230.8945  &  0.78  &  $  -4.80$  &    0.32  &  $ -0.55$  &  $  33.00$  &    0.29  &  $ -0.10$  &   Fairborn \\ 
2013 Feb 10  & 56333.6953  &  0.69  &  $  -3.40$  &    0.32  &  $ -0.29$  &  $  32.10$  &    0.29  &  $  0.18$  &   Fairborn \\ 
2013 May 22  & 56434.9688  &  0.55  &  $   0.20$  &    0.32  &  $ -0.52$  &  $  27.90$  &    0.29  &  $ -0.07$  &   Fairborn \\ 
 2013 Sep 06  & 56541.7578  &  0.57  &  $   0.30$  &    0.32  &  $  0.22$  &  $  28.90$  &    0.29  &  $  0.26$  &   Fairborn \\ 
2013 Nov 26  & 56622.8203  &  0.86  &  $  -2.10$  &    0.32  &  $  0.03$  &  $  31.00$  &    0.29  &  $  0.09$  &   Fairborn \\ 
2013 Dec 26  & 56652.5859  &  0.70  &  $  -3.60$  &    0.32  &  $ -0.17$  &  $  32.30$  &    0.29  &  $  0.05$  &   Fairborn \\ 
2014 May 28  & 56805.9492  &  0.04  &  $  64.20$  &    0.32  &  $ -0.17$  &  $ -37.80$  &    0.29  &  $ -0.17$  &   Fairborn \\ 
2014 Jun 26  & 56834.9609  &  0.86  &  $  -2.30$  &    0.32  &  $ -0.04$  &  $  30.80$  &    0.29  &  $ -0.25$  &   Fairborn \\ 
 2014 Oct 01  & 56931.7422  &  0.60  &  $  -0.80$  &    0.32  &  $ -0.05$  &  $  29.50$  &    0.29  &  $  0.00$  &   Fairborn \\ 
2014 Nov 16  & 56977.6133  &  0.89  &  $   1.20$  &    0.32  &  $  0.07$  &  $  27.70$  &    0.29  &  $  0.15$  &   Fairborn \\ 
 2015 Jun 07  & 57180.9258  &  0.64  &  $  -2.00$  &    0.32  &  $  0.04$  &  $  31.00$  &    0.29  &  $  0.18$  &   Fairborn \\ 
2015 Nov 10  & 57336.6836  &  0.05  &  $  61.70$  &    0.23  &  $  0.23$  &  $ -34.10$  &    0.27  &  $  0.54$  &   Fairborn \\ 
 2015 Dec 01  & 57357.6445  &  0.64  &  $  -1.81$  &    1.10  &  $  0.13$  &  $  30.03$  &    1.18  &  $ -0.69$  &      APO \\ 
2016 Feb 18  & 57436.6680  &  0.87  &  $  -1.20$  &    0.23  &  $  0.09$  &  $  30.20$  &    0.27  &  $  0.15$  &   Fairborn \\ 
 2016 Aug 07  & 57607.9570  &  0.72  &  $  -3.30$  &    0.23  &  $  0.34$  &  $  32.50$  &    0.27  &  $  0.03$  &   Fairborn \\ 
2016 Sep 14  & 57645.7852  &  0.79  &  $  -3.85$  &    1.08  &  $  0.39$  &  $  32.33$  &    1.19  &  $ -0.76$  &      APO \\ 
2016 Oct 14  & 57675.8672  &  0.64  &  $  -2.00$  &    0.23  &  $ -0.12$  &  $  30.90$  &    0.27  &  $  0.25$  &   Fairborn \\ 
2016 Oct 15  & 57676.6016  &  0.66  &  $  -2.89$  &    1.12  &  $ -0.49$  &  $  29.09$  &    1.25  &  $ -2.11$  &      APO \\ 
2016 Oct 21  & 57682.7852  &  0.83  &  $  -3.24$  &    1.09  &  $  0.34$  &  $  31.26$  &    1.20  &  $ -1.15$  &      APO \\ 
2016 Nov 18  & 57710.8008  &  0.62  &  $  -1.40$  &    0.23  &  $  0.14$  &  $  30.20$  &    0.27  &  $ -0.11$  &   Fairborn \\ 
2016 Nov 19  & 57711.6445  &  0.65  &  $  -2.23$  &    1.08  &  $ -0.06$  &  $  29.20$  &    1.22  &  $ -1.76$  &      APO \\ 
 2017 Jan 06  & 57759.7773  &  0.01  &  $  74.10$  &    0.23  &  $  0.05$  &  $ -47.80$  &    0.27  &  $ -0.19$  &   Fairborn \\ 
2017 May 17  & 57890.9727  &  0.72  &  $  -3.90$  &    0.23  &  $ -0.23$  &  $  32.70$  &    0.27  &  $  0.20$  &   Fairborn \\ 
2017 Jun 30  & 57934.9219  &  0.96  &  $  28.80$  &    0.23  &  $ -0.09$  &  $  -1.10$  &    0.27  &  $ -0.04$  &   Fairborn \\ 
2017 Aug 30  & 57995.9961  &  0.69  &  $  -2.80$  &    0.23  &  $  0.30$  &  $  31.80$  &    0.27  &  $ -0.12$  &   Fairborn \\ 
 2017 Sep 02  & 57998.7617  &  0.77  &  $  -4.04$  &    1.09  &  $  0.17$  &  $  32.09$  &    1.20  &  $ -0.96$  &      APO \\ 
 2017 Oct 04  & 58030.7109  &  0.67  &  $  -2.60$  &    0.23  &  $  0.09$  &  $  31.80$  &    0.27  &  $  0.30$  &   Fairborn \\ 
 2017 Nov 09  & 58066.6758  &  0.69  &  $  -3.00$  &    0.23  &  $  0.07$  &  $  32.20$  &    0.27  &  $  0.32$  &   Fairborn \\ 
2017 Dec 27  & 58114.6133  &  0.04  &  $  64.93$  &    0.53  &  $  1.05$  &  $ -35.67$  &    0.58  &  $  1.46$  &      APO \\ 
2018 Sep 27  & 58388.7539  &  0.79  &  $  -2.34$  &    1.16  &  $  1.88$  &  $  33.75$  &    1.30  &  $  0.68$  &      APO \\ 
2019 Jan 14  & 58497.5898  &  0.87  &  $  -0.59$  &    1.05  &  $  1.02$  &  $  30.19$  &    1.19  &  $ -0.18$  &      APO \\ 
2019 Jan 21  & 58504.7109  &  0.07  &  $  50.80$  &    0.23  &  $  0.19$  &  $ -23.20$  &    0.27  &  $  0.25$  &   Fairborn \\ 
2019 Jan 22  & 58505.5664  &  0.10  &  $  42.12$  &    0.52  &  $ -0.25$  &  $ -15.14$  &    0.58  &  $ -0.19$  &      APO \\ 
2019 Aug 18  & 58713.9609  &  0.99  &  $  58.81$  &    0.54  &  $  0.04$  &  $ -31.52$  &    0.60  &  $  0.34$  &      APO \\ 
2019 Sep 13  & 58739.9023  &  0.72  &  $  -2.79$  &    1.14  &  $  0.93$  &  $  32.03$  &    1.27  &  $ -0.51$  &      APO \\ 
2019 Oct 21  & 58777.8125  &  0.79  &  $  -3.21$  &    1.06  &  $  1.01$  &  $  32.85$  &    1.18  &  $ -0.22$  &      APO \\ 
\enddata  
\end{deluxetable*}

\begin{deluxetable*}{lccrrrrrrc}
\tablewidth{0pt}
\tabletypesize{\footnotesize}
\tablecaption{  Radial Velocity Measurements for HD~24546\label{rvtable24546}    }
\tablehead{ \colhead{UT Date} & \colhead{HJD-2,400,000} & \colhead{Orbital} & \colhead{$V_{r1}$} 
& \colhead{$\sigma_1$} & \colhead{$O-C$}  & \colhead{$V_{r2}$} & \colhead{$\sigma_2$} & \colhead{$O-C$} & \colhead{Source}
\\  
\colhead{} & \colhead{} & \colhead{Phase} & \colhead{(km~s$^{-1}$)} & \colhead{(km~s$^{-1}$)} 
& \colhead{(km~s$^{-1}$)} & \colhead{(km~s$^{-1}$)} & \colhead{(km~s$^{-1}$)} & \colhead{(km~s$^{-1}$)} & \colhead{}   }
\startdata 
2003 Nov 26 	& 52970.0430  &   0.41  & $ 45.70$  &  0.26  &  $ 0.59$  &    5.49  &  0.25  &  $ 0.08$  &    Fairborn  \\ 
2004 Oct 12 	& 53291.0273  &   0.96  & $-44.73$  &  0.26  &  $-0.21$  &   96.74  &  0.25  &  $ 0.15$  &    Fairborn  \\ 
2004 Nov 13 	& 53322.9648  &   0.01  & $-36.89$  &  0.26  &  $-0.02$  &   88.71  &  0.25  &  $-0.11$  &    Fairborn  \\ 
2004 Dec 13 	& 53352.9688  &   0.99  & $-52.89$  &  0.26  &  $ 0.10$  &  105.29  &  0.25  &  $ 0.08$  &    Fairborn  \\ 
2004 Dec 14 	& 53353.9219  &   0.02  & $-11.73$  &  0.26  &  $ 0.16$  &   63.33  &  0.25  &  $-0.07$  &    Fairborn  \\ 
2004 Dec 20 	& 53359.9219  &   0.22  & $ 48.01$  &  0.26  &  $ 0.29$  &    2.88  &  0.25  &  $ 0.12$  &    Fairborn  \\ 
2005 Apr 14 	& 53474.6875  &   0.99  & $-53.63$  &  0.26  &  $ 0.16$  &  106.05  &  0.25  &  $ 0.03$  &    Fairborn  \\ 
2005 Nov 11 	& 53686.0156  &   0.93  & $-28.00$  &  0.26  &  $ 0.01$  &   79.33  &  0.25  &  $-0.47$  &    Fairborn  \\ 
2006 Feb 13 	& 53779.8125  &   0.02  & $-23.95$  &  0.26  &  $-0.04$  &   75.90  &  0.25  &  $ 0.27$  &    Fairborn  \\ 
2006 Mar 26 	& 53820.6992  &   0.36  & $ 47.00$  &  0.26  &  $ 0.40$  &    4.29  &  0.25  &  $ 0.39$  &    Fairborn  \\ 
2006 Apr 21 	& 53846.6680  &   0.21  & $ 48.44$  &  0.26  &  $ 0.88$  &    3.22  &  0.25  &  $ 0.30$  &    Fairborn  \\ 
2016 Jan 26 	& 57413.6328  &   0.40  & $ 45.66$  &  0.98  &  $ 0.07$  &    5.28  &  0.86  &  $ 0.36$  &       APO  \\ 
2016 Oct 21 	& 57682.7812  &   0.24  & $ 47.43$  &  1.00  &  $-0.49$  &    2.35  &  0.86  &  $-0.21$  &       APO  \\ 
2016 Nov 19 	& 57711.6602  &   0.19  & $ 46.26$  &  0.96  &  $-0.59$  &    3.35  &  0.83  &  $-0.29$  &       APO  \\ 
2016 Dec 15 	& 57737.6445  &   0.04  & $  8.36$  &  1.01  &  $ 0.18$  &   43.09  &  0.93  &  $ 0.11$  &        APO  \\ 
2017 Jan 11 	& 57764.8633  &   0.94  & $-29.54$  &  0.36  &  $-0.65$  &   79.67  &  0.31  &  $-1.03$  &       APO  \\ 
2017 Dec 27 	& 58114.6445  &   0.43  & $ 45.12$  &  0.97  &  $ 0.48$  &    6.54  &  0.87  &  $ 0.65$   &       APO  \\ 
2018 Jan 28 	& 58146.8242  &   0.48  & $ 41.78$  &  0.96  &  $-0.70$  &    7.49  &  0.86  &  $-0.60$  &       APO  \\ 
2018 Sep 27 	& 58388.7695  &   0.43  & $ 44.47$  &  1.09  &  $ 0.02$  &    6.29  &  0.98  &  $ 0.19$  &       APO  \\ 
2018 Dec 24 	& 58476.6680  &   0.32  & $ 47.95$  &  0.99  &  $ 0.55$  &    3.99  &  0.84  &  $ 0.90$  &       APO  \\ 
2019 Jan 14 	& 58497.5977  &   0.01  & $-34.84$  &  0.36  &  $ 0.69$  &   87.64  &  0.31  &  $ 0.19$  &       APO  \\ 
2019 Jan 15 	& 58498.6328  &   0.04  & $  9.18$  &  1.12  &  $ 0.47$  &   42.24  &  1.02  &  $-0.21$  &       APO  \\ 
2019 Jan 19 	& 58502.8281  &   0.18  & $ 46.81$  &  0.97  &  $ 0.30$  &    4.62  &  0.83  &  $ 0.64$  &       APO  \\ 
2019 Jan 22 	& 58505.5391  &   0.27  & $ 47.19$  &  1.07  &  $-0.76$  &    1.82  &  0.94  &  $-0.70$  &       APO  \\ 
2019 Mar 24 	& 58566.5859  &   0.27  & $ 47.41$  &  1.00  &  $-0.51$  &    2.23  &  0.86  &  $-0.32$  &       APO  \\ 
2019 Sep 13 	& 58739.9375  &   0.97  & $-51.78$  &  0.37  &  $ 0.51$  &  105.06  &  0.32  &  $ 0.56$  &       APO  \\ 
2019 Sep 19 	& 58745.7422  &   0.16  & $ 45.26$  &  0.26  &  $-0.15$  &    4.95  &  0.25  &  $-0.16$  &    Fairborn  \\ 
2019 Oct 09  	& 58765.6914  &   0.82  & $ 13.70$  &  0.26  &  $-0.06$  &   36.93  &  0.25  &  $-0.37$  &    Fairborn  \\ 
2019 Oct 14 	& 58770.8164  &   0.98  & $-56.40$  &  0.37  &  $ 0.04$  &  108.78  &  0.32  &  $ 0.05$  &       APO  \\ 
2019 Oct 20 	& 58776.7461  &   0.18  & $ 46.36$  &  0.26  &  $-0.11$  &    3.81  &  0.25  &  $-0.22$  &    Fairborn  \\ 
2019 Oct 21 	& 58777.6445  &   0.21  & $ 47.30$  &  0.26  &  $-0.17$  &    2.72  &  0.25  &  $-0.29$  &    Fairborn  \\ 
2019 Oct 21 	& 58777.8789  &   0.22  & $ 48.35$  &  0.96  &  $ 0.72$  &    3.92  &  0.81  &  $ 1.07$  &       APO  \\ 
2019 Oct 25 	& 58781.6289  &   0.34  & $ 47.08$  &  0.26  &  $ 0.04$  &    3.11  &  0.25  &  $-0.34$  &    Fairborn  \\ 
2019 Oct 29 	& 58785.6875  &   0.47  & $ 43.08$  &  0.26  &  $ 0.12$  &    7.40  &  0.25  &  $-0.20$  &    Fairborn  \\ 
2019 Oct 30 	& 58786.6250  &   0.50  & $ 41.42$  &  0.26  &  $-0.24$  &    8.50  &  0.25  &  $-0.43$  &    Fairborn  \\ 
2019 Oct 31 	& 58787.6250  &   0.54  & $ 40.16$  &  0.26  &  $ 0.05$  &   10.57  &  0.25  &  $ 0.07$  &    Fairborn  \\ 
2019 Nov 01  	& 58788.6250  &   0.57  & $ 38.44$  &  0.26  &  $ 0.06$  &   12.12  &  0.25  &  $-0.14$  &    Fairborn  \\ 
2019 Nov 09  	& 58796.8359  &   0.84  & $  8.89$  &  0.26  &  $-0.10$  &   41.97  &  0.25  &  $-0.19$  &    Fairborn  \\ 
2019 Nov 13 	& 58800.8359  &   0.97  & $-52.84$  &  0.26  &  $-0.17$  &  104.75  &  0.25  &  $-0.14$  &    Fairborn  \\ 
2019 Nov 14 	& 58801.8359  &   0.00  & $-42.44$  &  0.26  &  $-0.05$  &   94.60  &  0.25  &  $ 0.17$  &    Fairborn  \\ 
2019 Nov 14 	& 58801.8906  &   0.00  & $-39.35$  &  0.34  &  $ 0.63$  &   92.01  &  0.30  &  $ 0.03$  &       APO  \\ 
2019 Nov 15 	& 58802.8359  &   0.04  & $  2.28$  &  0.26  &  $-0.32$  &   48.49  &  0.25  &  $-0.17$  &    Fairborn  \\ 
2019 Nov 17 	& 58804.8359  &   0.10  & $ 37.63$  &  0.26  &  $-0.11$  &   12.91  &  0.25  &  $-0.00$  &    Fairborn  \\ 
2019 Nov 18 	& 58805.8359  &   0.13  & $ 43.19$  &  0.26  &  $ 0.10$  &    7.34  &  0.25  &  $-0.13$  &    Fairborn  \\ 
2019 Nov 23 	& 58810.8086  &   0.30  & $ 47.79$  &  0.26  &  $ 0.07$  &    2.73  &  0.25  &  $-0.03$  &    Fairborn  \\ 
2020 Jan 12 	& 58860.6758  &   0.94  & $-28.98$  &  0.34  &  $ 0.20$  &   80.93  &  0.30  &  $-0.07$  &       APO  \\ 
\enddata  
\end{deluxetable*}

\clearpage

\begin{figure*}[h!]
\centering
\includegraphics[width=0.49\textwidth, trim = 30 8 50 10, clip]{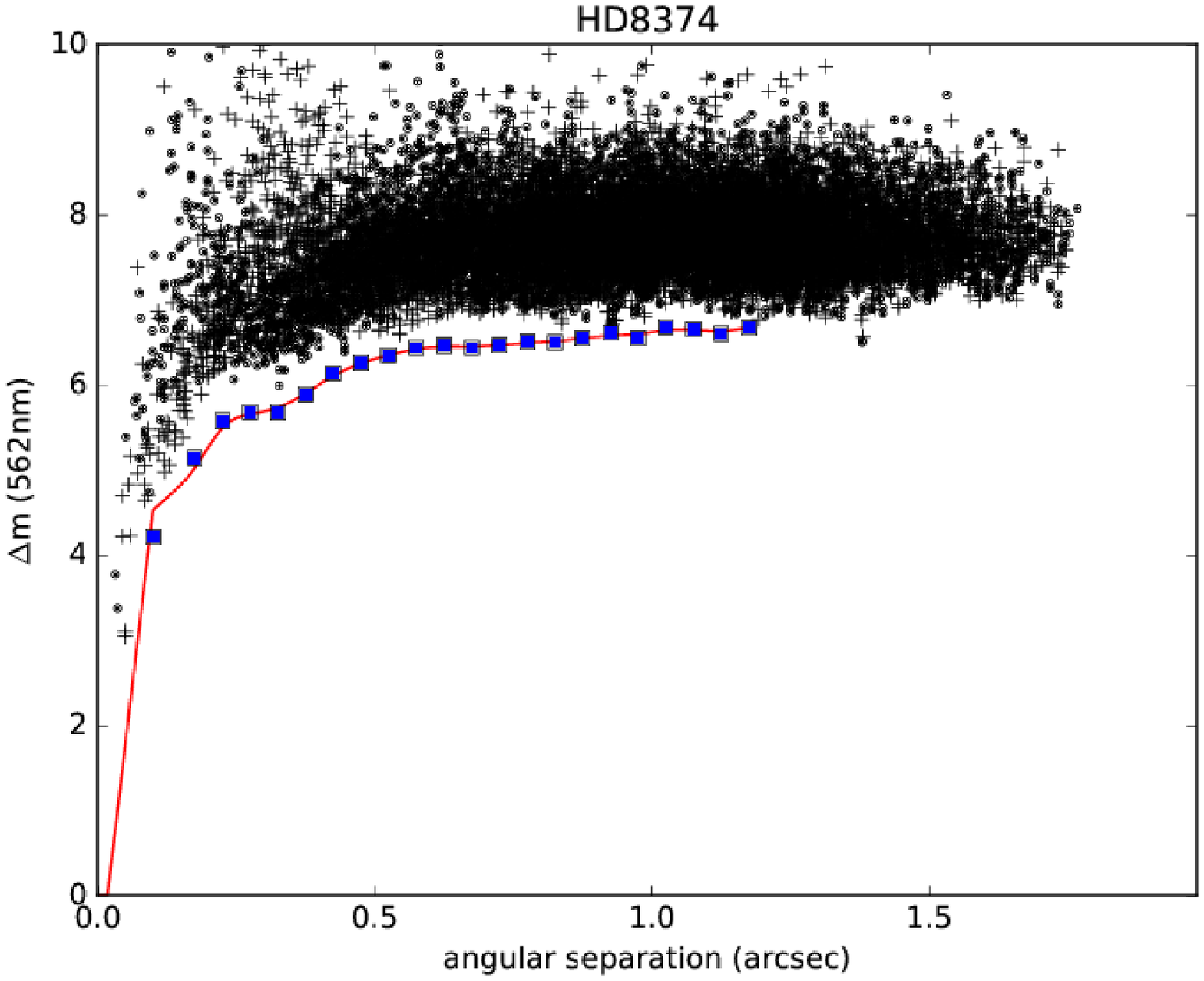} \hfill
\includegraphics[width=0.49\textwidth, trim = 30 8 50 10, clip]{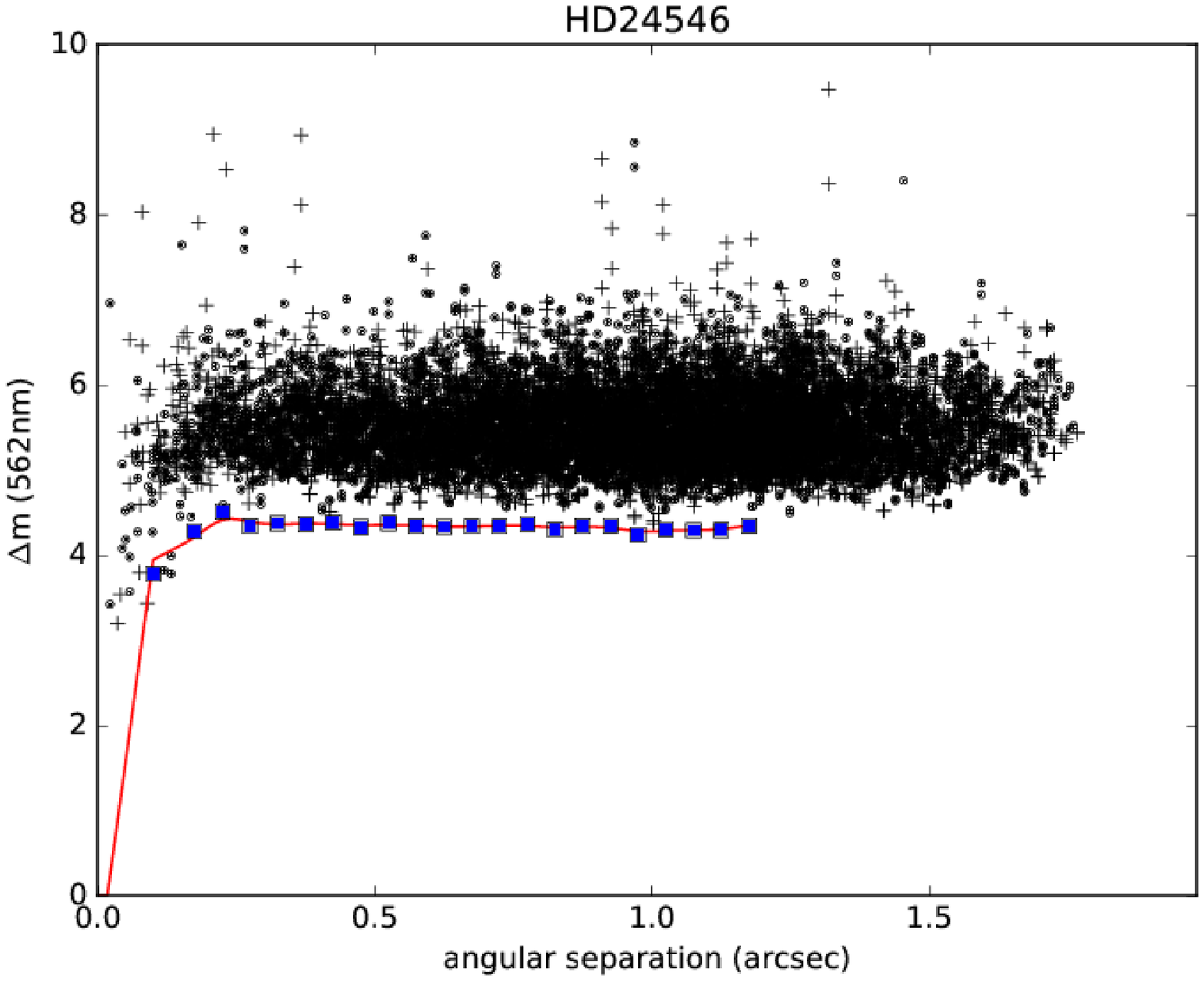}
\caption{Background sensitivity curves as a function of radius from the center for the reconstructed speckle images. The black points represent the local maxima (crosses) and minima (dots). The blue squares mark the $5\sigma$ background sensitivity limit within $0.05\arcsec$ bins, and the red line corresponds to a spline fit \citep{horch17}. No points fall below the contrast limits, therefore no tertiary companions were detected for HD~8374 or HD~24546. \label{speckle}}
\end{figure*}

\section{Interferometry}\label{section:inter}

\subsection{`Alopeke Observations \label{alopeke}  }
We observed these stars on 2018 Oct 23 with the `Alopeke speckle camera \citep{scott18} on the 8.0~m Gemini-North telescope\footnote{\href{https://www.gemini.edu/sciops/instruments/alopeke-zorro/}{https://www.gemini.edu/sciops/instruments/alopeke-zorro/}} in order to search for any unresolved companions outside of CHARA's limits that would bias our results.  `Alopeke takes a series of 1000 60~ms exposures simultaneously in the 562~nm and 716~nm bands, and the data were reduced with the speckle team's pipeline \citep{howell11}.  We searched for companions within 1.5\arcsec\ using the method given by \citet{horch17}, but found no hidden companions down to the detection limits of $\Delta m < 6$ mag for HD~8374 and $\Delta m < 4$ mag for HD~24546. Plots of the detection limits for the 562~nm filter are shown in Figure~\ref{speckle}.

\subsection{CHARA Array Observations} 
We observed HD~8374 and HD~24546 with the CHARA Array from 2017--2019. CHARA has six 1.0~m telescopes arranged in a Y-shape with baselines ranging from 34--331~m. We used the CLIMB beam combiner to combine the $K'$-band light from three telescopes at a time. Our observations are listed in Tables~\ref{charalog8374} and \ref{charalog24546}, with the UT and heliocentric Julian dates, the telescope combination used, and the number of visibilities and closure phases measured.  The CLIMB data were reduced with the pipeline developed by John D. Monnier, with the general method described in \citet{monnier11} and extended to three beams \citep[e.g., ][]{kluska18}, resulting in squared visibilities ($V^2$) for each baseline and closure phases (CP) for each closed triangle. We corrected for any instrumental and atmospheric effects on the observed visibilities using observations of stars with known angular diameters taken before and after the target to complete an observation ``bracket". The calibrator stars for HD~8374 were HD~8774 and HD~9616, which have $K$-band uniform disk angular diameters of $0.38\pm0.01$~mas and $0.38\pm0.01$~mas from SearchCal \citep{searchcal}. The calibrators for HD~24546 were HD~25642 and HD~27084, which have $K$-band angular diameters of $0.48\pm0.03$~mas and $0.40\pm0.04$~mas.

\begin{deluxetable*}{lccccc}	
\tablewidth{0pt}
\tabletypesize{\footnotesize}
\tablecaption{Observing Log for HD 8374  \label{charalog8374}}
\tablehead{ \colhead{UT Date} & \colhead{HJD-2,400,000} & \colhead{Source}  & \colhead{Telescope} 
& \colhead{Number}  & \colhead{Number}   \\
\colhead{ } & \colhead{ } & \colhead{} & \colhead{Configuration} & \colhead{of $V^2$}  & \colhead{of CP} }
\startdata  
1999 Aug 11   &   51401.975   &  PTI   &  N-S       &   7   &  \nodata  \\
1999 Aug 23   &   51413.988   &  PTI   &  N-S       &   4   &  \nodata  \\
1999 Sep 03   &   51424.932   &  PTI   &  N-S       &   5   &  \nodata  \\
1999 Oct 13   &   51464.836   &  PTI   &  N-S       &   4   &  \nodata  \\
1999 Oct 20   &   51471.808   &  PTI   &  N-S       &   5   &  \nodata  \\
1999 Oct 26   &   51477.806   &  PTI   &  N-S       &   6   &  \nodata  \\
2001 Aug 27   &   52148.972   &  PTI   &  S-W       &   6   &  \nodata  \\
2001 Aug 30   &   52151.973   &  PTI   &  S-W       &   6   &  \nodata  \\
2001 Sep 25   &   52177.884   &  PTI   &  S-W       &   6   &  \nodata  \\
2005 Aug 12   &   53594.977   &  PTI   &  S-W       &   6   &  \nodata  \\
2005 Aug 19   &   53601.963   &  PTI   &  S-W       &   8   &  \nodata  \\
2005 Oct 23   &   53666.784   &  PTI   &  S-W       &   7   &  \nodata  \\
2005 Nov 14   &   53688.717   &  PTI   &  N-S       &  12   &  \nodata  \\
2005 Nov 15   &   53689.679   &  PTI   &  N-W       &   5   &  \nodata  \\
2017 Sep 07   &   58003.862   &  CHARA &  E1-W1-W2  &   12  &   4       \\
2017 Nov 30   &   58087.624   &  CHARA &  S1-W1-E1  &   9   &   3       \\
2018 Aug 15   &   58345.929   &  CHARA &  S1-W1-E1  &   12  &   4       \\
2018 Aug 16   &   58346.811   &  CHARA &  S1-W1-E1  &   21  &   7       \\
2018 Aug 17   &   58347.889   &  CHARA &  S1-W1-E1  &   18  &   6       \\
2018 Sep 03   &   58364.787   &  CHARA &  S1-W1-E1  &   12  &   4       \\
2018 Sep 04   &   58365.783   &  CHARA &  S1-W1-E1  &   9   &   3       \\
2019 Sep 16   &   58742.869   &  CHARA &  S1-W1-E1  &   12  &   4       \\
2019 Sep 17   &   58743.836   &  CHARA &  S1-W1-E1  &   12  &   4       \\
2019 Sep 18   &   58744.818   &  CHARA &  S1-W1-E1  &   12  &   4       \\
\enddata                          
\end{deluxetable*}

\subsection{PTI Observations}  
We also observed HD~8374 using the Palomar Testbed Interferometer \citep[PTI,][]{pti1} from 1999--2005 as listed in Table~\ref{charalog8374}. PTI had three 40 cm telescopes with separations of 87--110~m and combined the near-infrared light from two telescopes at a time.  The single baseline measured only squared visibilities, because at least three baselines are needed to measure closure phases. All of the observations were taken in $K$-band, except on 1999 Sep 03 which were taken in $H$-band. These data were reduced using the standard PTI reduction pipeline \citep{pti2} and calibrated using the software provided by NExScI\footnote{\href{http://nexsci.caltech.edu/software/}{http://nexsci.caltech.edu/software/}}. The observed calibrator stars were HD~6920, HD~7034, HD~7964, and HD~11007 with $K$-band angular diameters of $0.58\pm0.02$~mas \citep{vanbelle08}, $0.51\pm0.02$~mas \citep{boden06}, $0.42\pm0.04$~mas  \citep{boden06}, and $0.45\pm0.10$~mas \citep{konacki04}, respectively.

\begin{deluxetable*}{lccccc}	
\tablewidth{0pt}
\tabletypesize{\footnotesize}
\tablecaption{Observing Log for HD 24546 \label{charalog24546}}
\tablehead{ \colhead{UT Date} & \colhead{HJD-2,400,000} & \colhead{Source}  & \colhead{Telescope} 
& \colhead{Number}  & \colhead{Number}   \\
\colhead{ } & \colhead{ } & \colhead{} & \colhead{Configuration} & \colhead{of $V^2$}  & \colhead{of CP} }
\startdata 
2017 Sep 07   &   58003.984   & CHARA &   E1-W1-W2    &   12	&   4	  \\
2017 Sep 08   &   58004.982   & CHARA &   S1-W1-E1     &   24	&   8	  \\
2017 Oct 11   &   58037.975   & CHARA &   S1-W1-E1      &   21	&   7	  \\
2017 Nov 30   &   58087.753   & CHARA &   S1-W1-E1     &   15 	&   5	  \\
2018 Aug 17   &   58348.009   & CHARA &   S1-W1-E1     &   9	   	&   3	  \\	
2019 Sep 17   &   58743.993   & CHARA &   S1-W1-E1     &   18	&   6	  \\
2019 Sep 18   &   58744.912   & CHARA &   S1-W1-E1     &   12	&   4	  \\
2019 Dec 20   &   58837.781   &   CHARA & S1-W1-E1     &   18	&   6	   \\
2019 Dec 21   &   58838.653   &   CHARA & S1-W1-E1     &   18	&   6	    \\                          
\enddata                          
\end{deluxetable*}

\subsection{Binary Positions}  
We measured the relative positions from the interferometric visibilities and closure phases using the method\footnote{\href{http://www.chara.gsu.edu/analysis-software/binary-grid-search}{http://www.chara.gsu.edu/analysis-software/binary-grid-search}} of \citet{schaefer16} as described in \citet{lester19a}.  Briefly, we searched across a grid of separations in right ascension and declination for the best-fit relative position. At each grid point, we compared the observed $V^2$ and CP to model values to fit for the flux ratio and calculate the \chisq value. We then searched a small area around the best-fit position to fit an ellipse to the contour marking $\chi^2 \le \chi^2_{min}+1$, which determines the major axis, minor axis, and position angle of the error ellipse.  Because the orbital periods of these systems are quite long, any orbital motion within a single night is typically within the error ellipses.  The best-fit relative positions, error ellipse parameters, and flux ratio estimates for each night are listed in Tables~\ref{relpos8374} and \ref{relpos24546}.  The average $K'$-band flux ratio from the CHARA observations is $0.79\pm0.14$ for HD~8374 and $0.92\pm0.17$ for HD~24546, where the uncertainty corresponds to the standard deviation from all nights.   

The PTI observations of HD~8374 used only one baseline per night, so only one vector component of the separation could be measured. This resulted in multiple solutions within the $1\sigma$ \chisq limit, especially on nights with fewer than five $V^2$ points. Without closure phases to measure the flux asymmetry, each solution is also reflected across the origin. In order to break these ambiguities, one could either fit the visual orbit directly to the visibilities \citep[e.g.,][]{boden99, heminiak12}, or use the 3-telescope observations as a reference. We opted for the latter method and chose the PTI solutions most consistent with a preliminary visual orbit from the CHARA observations.

\begin{deluxetable*}{lccrrrrrc}
\tablewidth{0pt}
\tabletypesize{\footnotesize}
\tablecaption{Relative Positions for HD 8374 \vspace{6pt} \label{relpos8374}}
\tablehead{ \colhead{UT Date} & \colhead{HJD-2,400,000} & \colhead{Orbital} & \colhead{$\rho$} 
& \colhead{$\theta$} & \colhead{$\sigma_{maj}$} & \colhead{$\sigma_{min}$} & \colhead{$\phi$} & \colhead{$f_2/f_1$} \\
\colhead{ } & \colhead{ } & \colhead{Phase} & \colhead{(mas)} & \colhead{(deg)} &  \colhead{(mas)}  
&  \colhead{(mas)}  &  \colhead{(deg)}  &  \colhead{ }  }
\startdata 
1999 Aug 11  &  51401.9750  &   0.25 &  5.632 &   31.62 &   0.258 &   0.062 &    75.0 &  0.59 \\
1999 Aug 23  &  51413.9880  &   0.59 &  7.807 &  354.79 &   0.604 &   0.080 &    99.1 &  0.45 \\
1999 Sep 03  &  51424.9322  &   0.90 &  3.784 &  310.03 &   0.307 &   0.043 &   105.7 &  0.46 \\
1999 Oct 13  &  51464.8361  &   0.03 &  1.927 &  140.55 &   1.258 &   0.141 &   110.2 &  0.43 \\
1999 Oct 20  &  51471.8083  &   0.23 &  5.248 &   37.21 &   0.986 &   0.185 &   111.0 &  1.00 \\
1999 Oct 26  &  51477.8062  &   0.40 &  7.160 &   10.10 &   0.873 &   0.200 &   102.2 &  1.00 \\
2001 Aug 27  &  52148.9729  &   0.37 &  7.307 &   15.29 &   0.754 &   0.061 &   154.4 &  0.88 \\
2001 Aug 30  &  52151.9740  &   0.46 &  7.656 &    7.86 &   1.702 &   0.070 &   149.3 &  0.64 \\
2001 Sep 25  &  52177.8845  &   0.19 &  4.608 &   45.12 &   0.356 &   0.109 &   131.6 &  0.50 \\
2005 Aug 12  &  53594.9778  &   0.26 &  5.673 &   32.72 &   0.738 &   0.068 &   163.6 &  0.44 \\
2005 Aug 19  &  53601.9639  &   0.46 &  7.716 &    8.00 &   0.377 &   0.025 &   166.2 &  0.83 \\
2005 Oct 23  &  53666.7841  &   0.29 &  6.488 &   23.88 &   0.420 &   0.021 &   160.5 &  0.56 \\
2005 Nov 14  &  53688.7174  &   0.91 &  3.259 &  308.58 &   0.203 &   0.023 &   119.6 &  0.97 \\
2005 Nov 15  &  53689.6795  &   0.94 &  2.562 &  296.05 &   1.016 &   0.111 &    89.9 &  0.95 \\  
2017 Sep 07  &  58003.8658  &   0.91 &  3.218 &  302.72 &   0.454 &   0.028 &   170.9 &  0.75 \\
2017 Nov 30  &  58087.6281  &   0.28 &  5.921 &   28.32 &   0.087 &   0.037 &    53.8 &  0.68  \\
2018 Aug 15  &  58345.9310  &   0.59 &  7.827 &  355.95 &   0.390 &   0.129 &   154.5 &  0.72  \\
2018 Aug 16  &  58346.8129  &   0.61 &  7.531 &  354.32 &   0.209 &   0.148 &   116.8 &  0.65  \\
2018 Aug 17  &  58347.8908  &   0.64 &  7.667 &  351.17 &   0.133 &   0.136 &   180.0 &  0.87  \\
2018 Sep 03  &  58364.7905  &   0.12 &  3.450 &   69.81 &   0.022 &   0.019 &     4.0 &  0.79  \\
2018 Sep 04  &  58365.7858  &   0.15 &  3.943 &   57.84 &   0.057 &   0.025 &   136.3 &  0.98  \\
2019 Sep 16  &  58742.8731  &   0.81 &  5.905 &  332.33 &   0.201 &   0.201 &   131.9 &  0.65  \\
2019 Sep 17  &  58743.8404  &   0.84 &  5.347 &  328.28 &   0.049 &   0.018 &   166.5 &  0.98  \\
2019 Sep 18  &  58744.8179  &   0.86 &  4.566 &  322.29 &   0.238 &   0.110 &   134.6 &  1.01  \\
\enddata  
\end{deluxetable*}

\begin{deluxetable*}{lccrrrrrc}
\tablewidth{0pt}
\tabletypesize{\footnotesize}
\tablecaption{Relative Positions for HD 24546  \vspace{6pt} \label{relpos24546}}
\tablehead{ \colhead{UT Date} & \colhead{HJD-2,400,000} & \colhead{Orbital} & \colhead{$\rho$} 
& \colhead{$\theta$} & \colhead{$\sigma_{maj}$} & \colhead{$\sigma_{min}$} & \colhead{$\phi$} & \colhead{$f_2/f_1$} \\
\colhead{ } & \colhead{ } & \colhead{Phase} & \colhead{(mas)} & \colhead{(deg)} &  \colhead{(mas)}  
&  \colhead{(mas)}  &  \colhead{(deg)}  &  \colhead{ }  }
\startdata 
2017 Sep 07   &  58003.9840  &  0.80  &  5.385 &  200.28  &   0.323 &  0.040  &  169.6  & 0.87 \\
2017 Sep 08   &  58004.9818  &  0.83  &  4.663 &  209.18  &   0.083 &  0.064  &   69.7   & 0.90 \\
2017 Oct 11    &  58037.9752  &  0.91  &  2.983 &  258.48  &   0.139 &  0.078  &   55.6   & 0.69 \\
2017 Nov 30   &  58087.7533  &  0.55  &  9.733 &  168.72  &   0.328 &  0.089  &    3.5    & 0.79 \\
2018 Aug 17   &  58348.0094  &  0.10  &  4.635 &  125.20  &   0.114 &  0.087  &   73.9    & 0.96 \\
2019 Sep 17   &  58743.9935  &  0.11  &  5.082 &  128.84  &   0.108 &  0.073  &  140.6   & 0.65 \\
2019 Sep 18   &  58744.9119  &  0.14  &  6.167 &  135.93  &   0.088 &  0.088  &   15.5    & 1.00 \\
2019 Dec 20   &  58837.7811  &  0.19  &  7.961 &  143.18   &  0.098 &  0.059  & 145.65  & 0.50 \\
2019 Dec 21   &  58838.6529  &  0.22  &  8.549 &  146.10   &  0.164 &  0.073  & 129.71  & 0.66 \\
\enddata  
\end{deluxetable*} 

\break

\subsection{Combined Visual + Spectroscopic Solution}\label{vbsbfit}
\begin{figure*}[h!]
\centering
\includegraphics[width=0.49\textwidth]{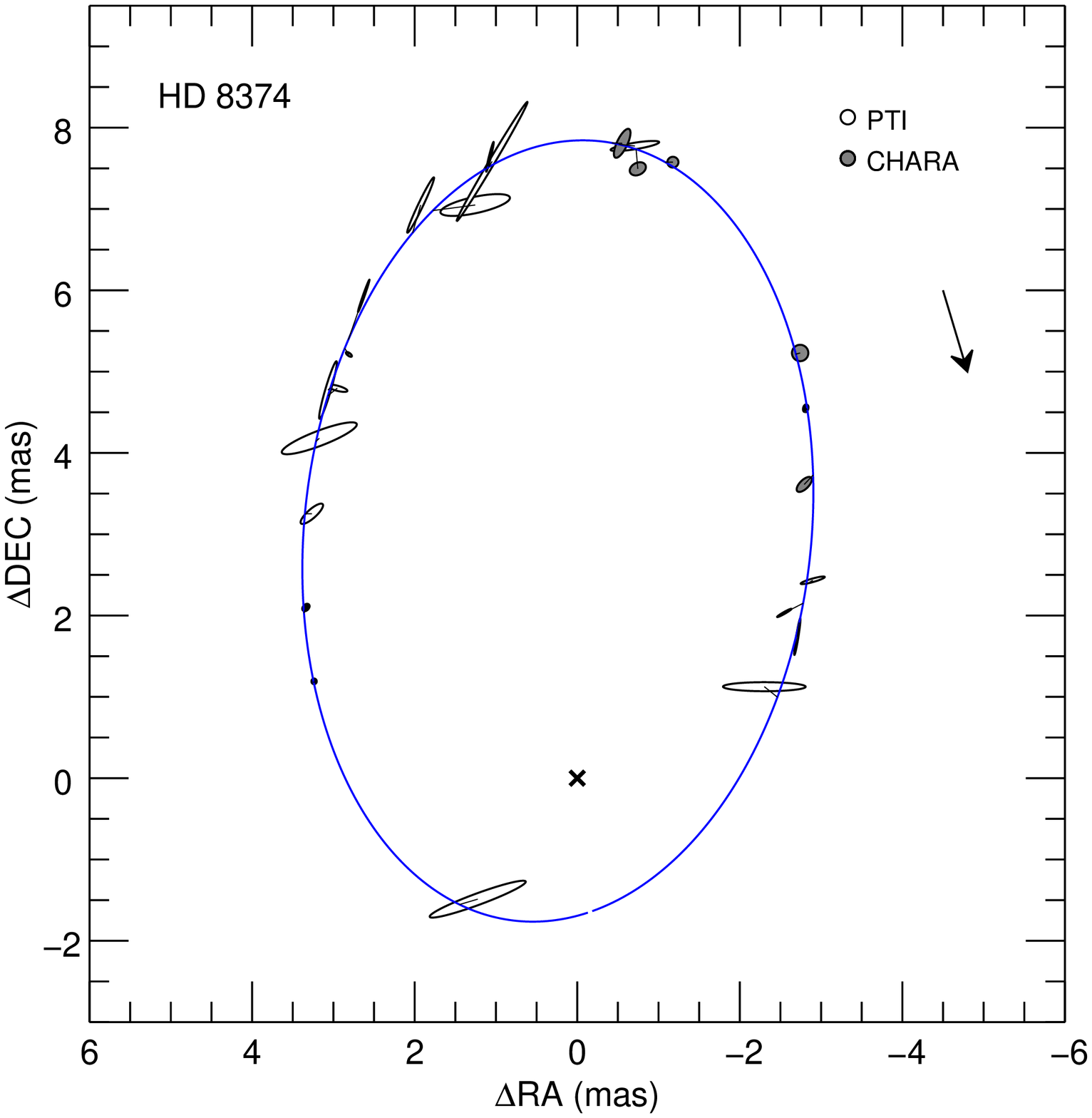} \hfill
\includegraphics[width=0.49\textwidth]{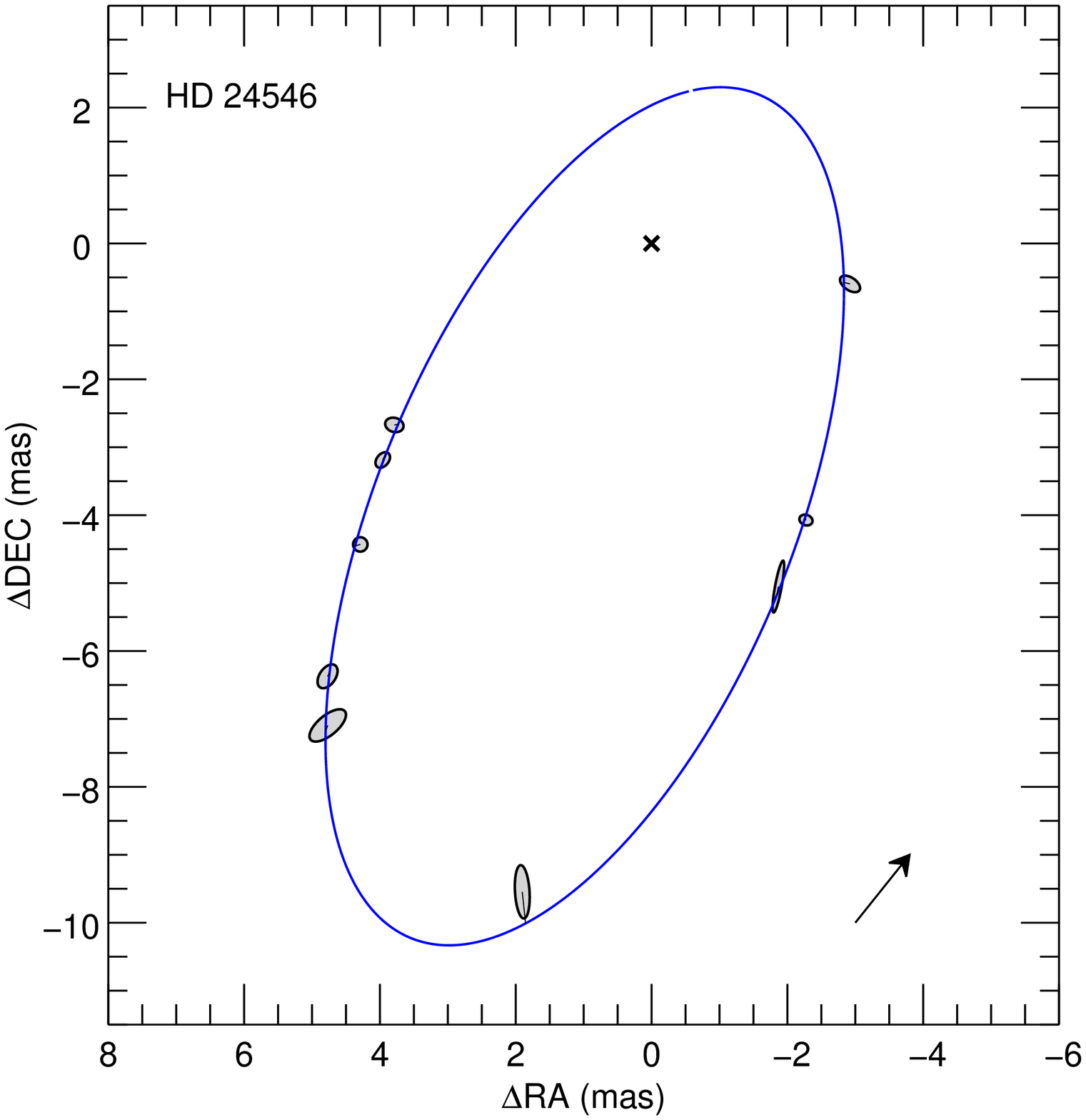} 
\caption{Visual orbit of HD~8374 (left) and HD~24546 (right). The primary star is located at the origin (black cross), and the relative positions of the secondary star are marked by the gray, filled ellipses (CHARA) and the open ellipses (PTI) corresponding to the size of the error ellipses. Several of the PTI data points have a very large axis ratio and appear as line segments. The solid blue curve represents the best-fit model visual orbit, and a thin black line connects each observed and model position. The arrow indicates the direction of orbital motion. \label{vborbit}}
\end{figure*}

\begin{figure*}[h!]
\centering
\includegraphics[width=0.49\textwidth]{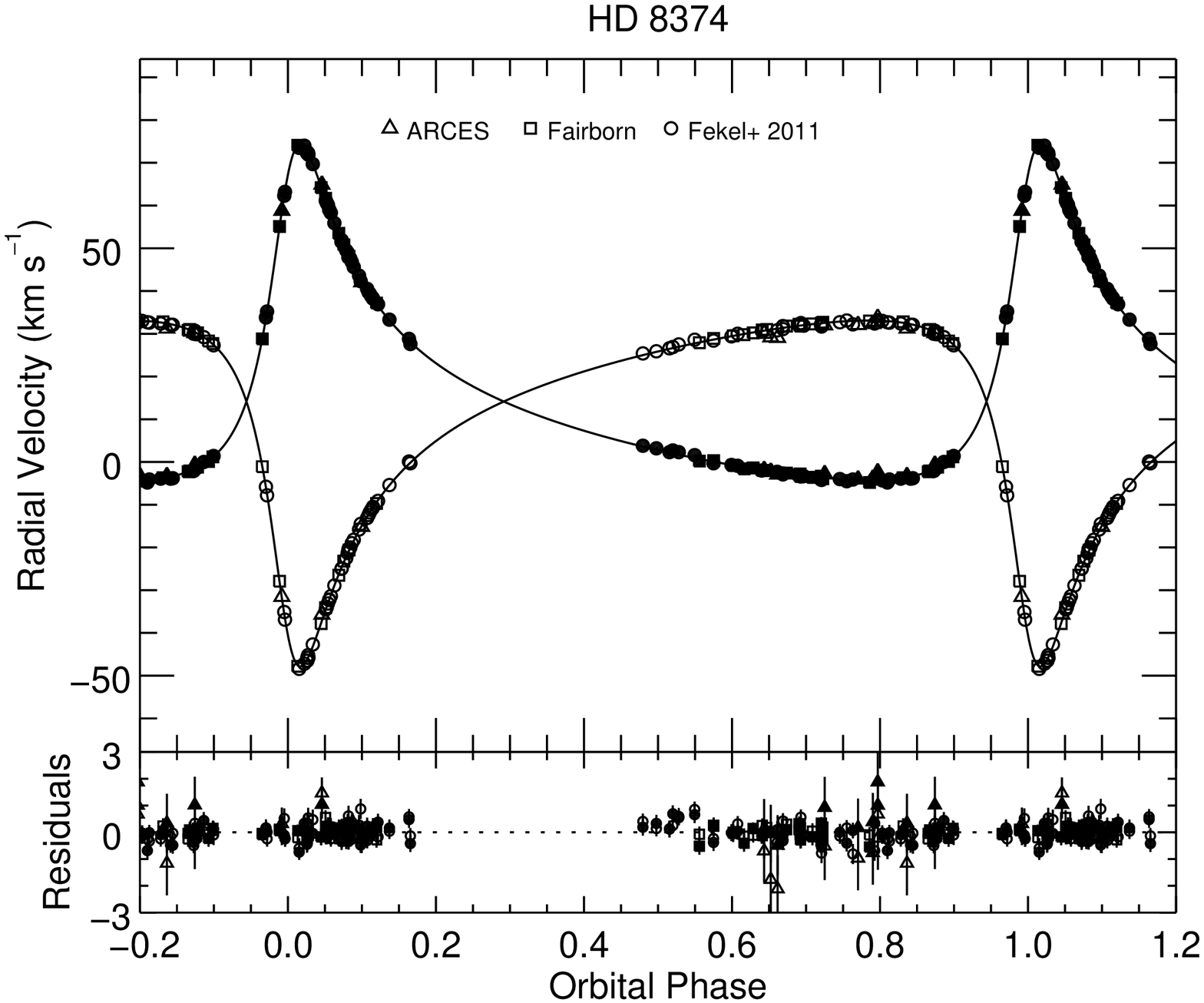}\hfill
\includegraphics[width=0.49\textwidth]{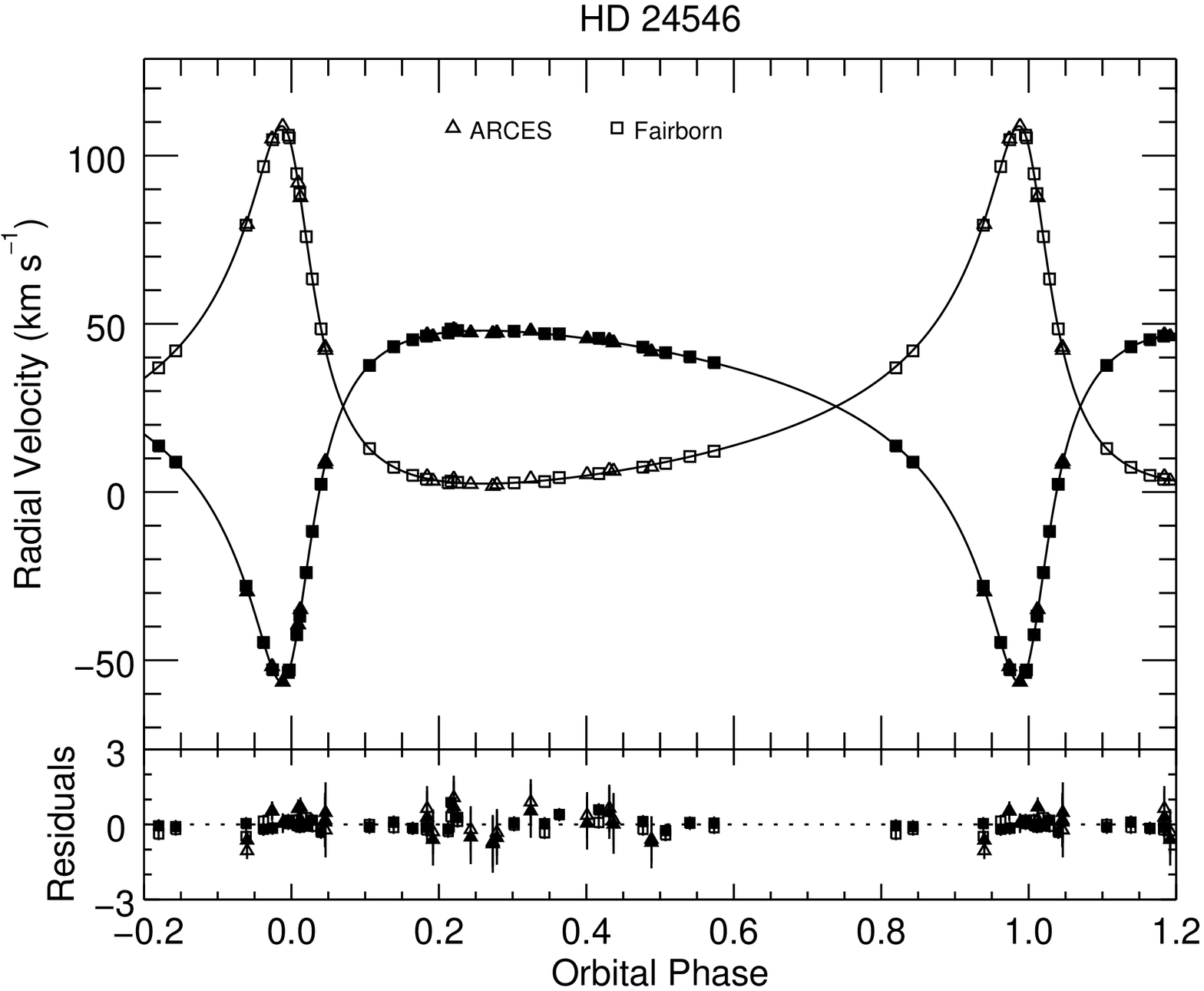}
\caption{Radial velocity curve of HD~8374 (left) and HD~24546 (right). The observed data for the primary and secondary star are shown with the filled and open points, respectively. The model curves are shown with the solid lines, and the residuals to the fit are shown in the bottom panels.  \label{rvcurve}}
\end{figure*}

We determined the final orbital solution by simultaneously fitting the interferometric and spectroscopic data using the method of \citet{schaefer16}. The full set of orbital parameters includes the orbital period ($P$), epoch of periastron ($T$), eccentricity ($e$), longitude of periastron of the primary star ($\omega_1$), the inclination ($i$), the angular semi-major axis ($a$), the longitude of the ascending node ($\Omega$), the systemic velocity ($\gamma$), and the velocity semi-amplitudes ($K_1$, $K_2$). Table~\ref{orbpar} lists the best-fit orbital solutions for HD~8374 and HD~24546. The visual orbits are shown in Figure~\ref{vborbit}, and the spectroscopic orbits are shown in Figure~\ref{rvcurve}. To determine the uncertainty of each orbital parameter, we performed a Monte Carlo error analysis in which we varied each data point within its Gaussian uncertainty and refit for the orbital solution. We then made a histogram of the best-fit parameters from $10^5$ iterations and fit Gaussians to each distribution to determine the $1\sigma$ uncertainties in each parameter (listed in Table~\ref{orbpar}).

\begin{deluxetable}{lcc}
\tablewidth{0pt}
\tabletypesize{\footnotesize}
\tablecaption{Orbital Parameters from VB+SB2 Solution \vspace{6pt} \label{orbpar}}
\tablehead{
\colhead{Parameter} &\colhead{HD~8374} & \colhead{HD~24546} 
}
\startdata	
$P$ (d)			& $  35.36836 \pm 0.00005 $ 	&  $    30.43885 \pm0.00002$  \\ 
$T$ (HJD-2400000)	& $ 54293.208 \pm 0.004   $ 	&  $ 57340.551 \pm 0.003 $  \\ 
$e$				& $    0.6476 \pm 0.0005  $ 	&  $     0.6421 \pm 0.0006 $  \\ 
$\omega_1$ (deg)	& $    325.18 \pm 0.10    $ 	&  $   207.71 \pm 0.11 $  \\ 
$i$ (deg)			& $  140.64 \pm 0.45  $ 		&  $    56.76 \pm 0.45 $  \\ 
$a$ (mas)			& $    5.05 \pm 0.02  $ 		&  $     6.99 \pm 0.06 $  \\ 
$\Omega$ (deg)	& $  336.2 \pm 0.1  $ 		&  $   150.2 \pm 0.3$  \\ 
$\gamma$ (\kms) 	& $   14.14 \pm 0.02  $ 		&  $    25.43 \pm 0.04 $  \\ 
$K_1$ (\kms)		& $   39.27 \pm 0.05  $ 		&  $    52.24 \pm 0.06 $  \\ 
$K_2$ (\kms)		& $   40.47 \pm 0.05  $ 		&  $    53.15 \pm 0.06 $  \\  
\enddata                   
\end{deluxetable}

\begin{figure*}
\centering
\includegraphics[width=0.49\textwidth]{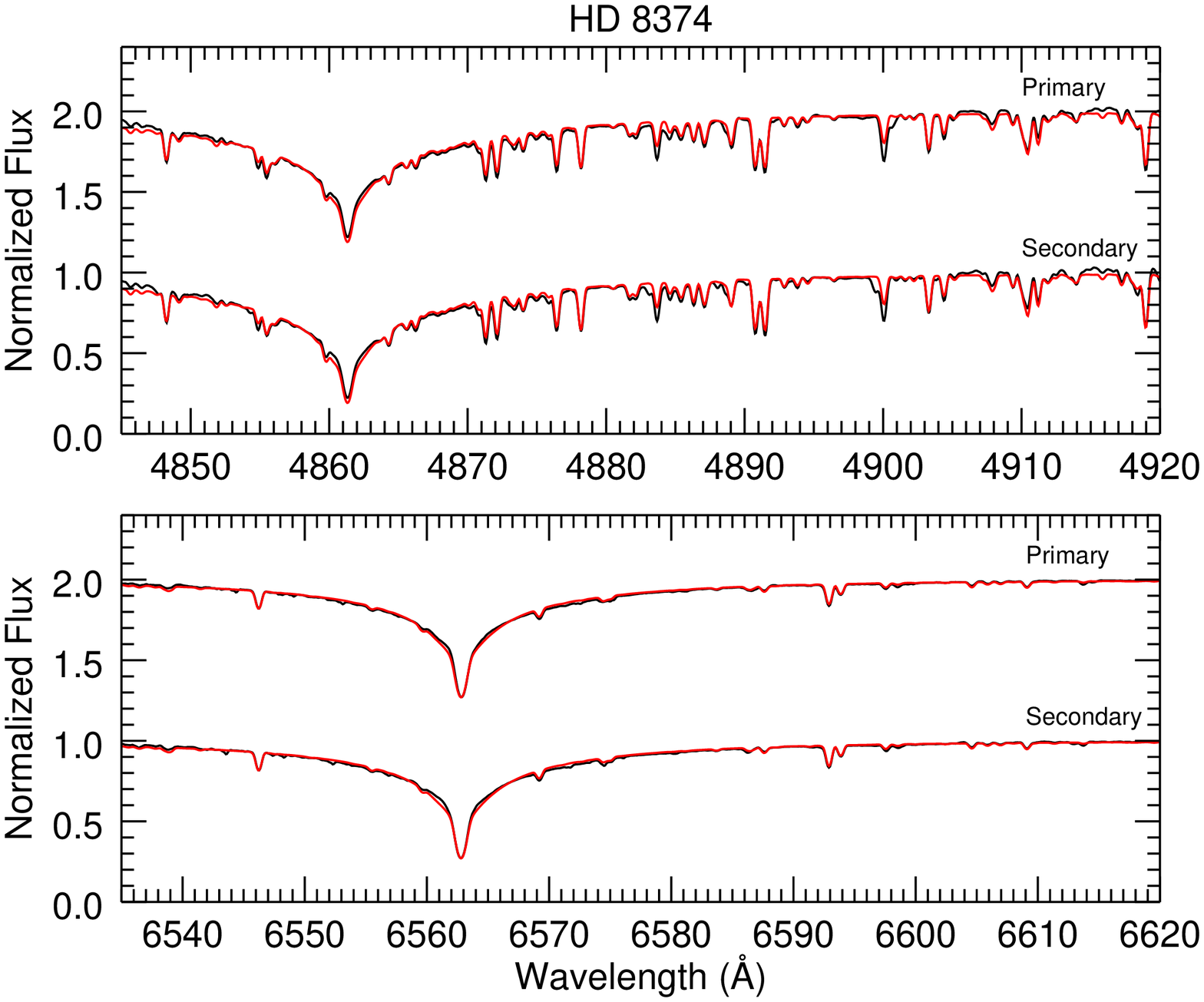} \hfill
\includegraphics[width=0.49\textwidth]{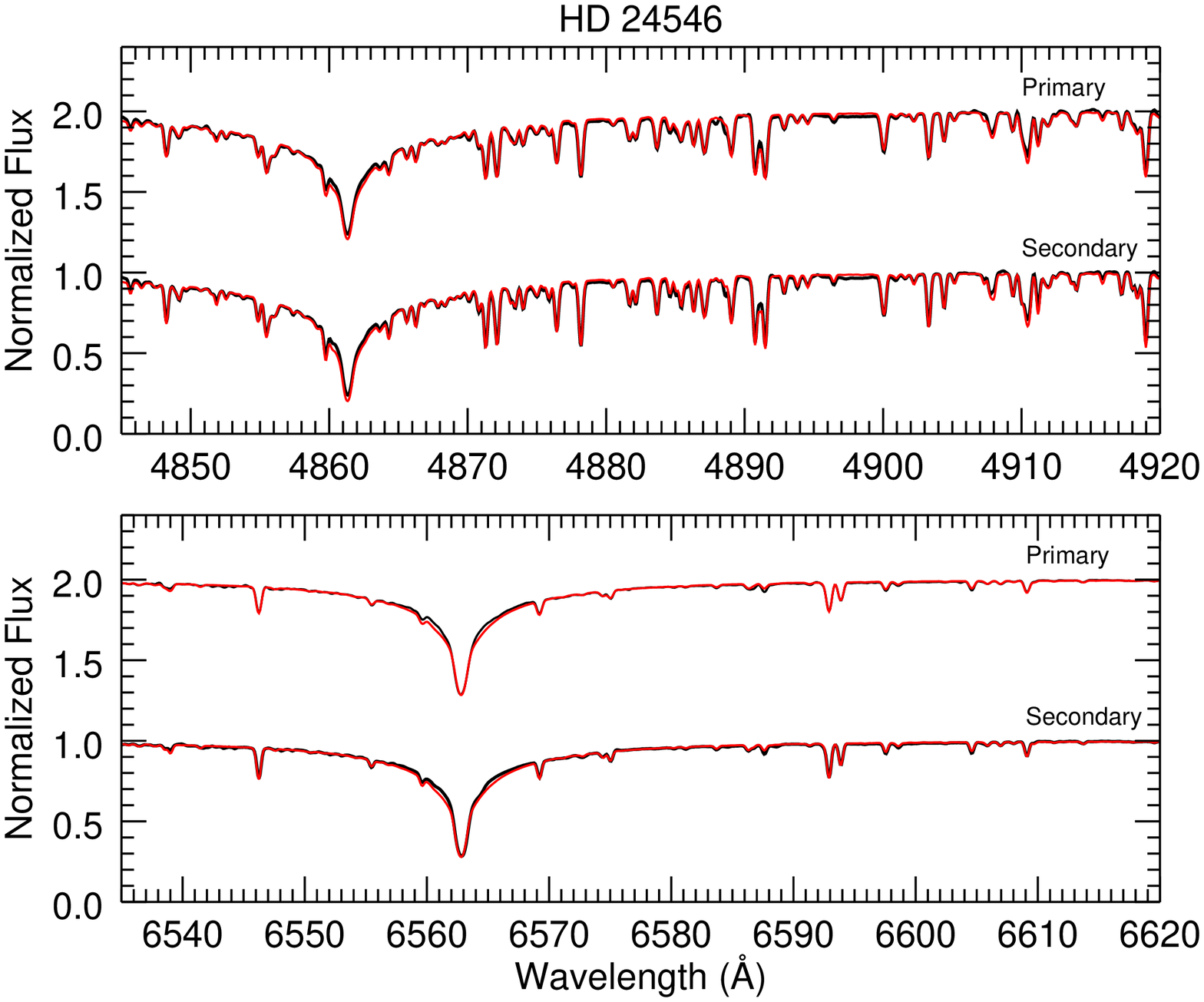} \caption{Reconstructed spectra of HD~8374 (left) and HD~24546 (right) around H$\beta$ and H$\alpha$. The reconstructed spectra are shown in black, and the best-fit model spectra are overplotted in red. \label{recspec}}
\end{figure*}

\begin{figure*}
\centering
\includegraphics[width=0.49\textwidth]{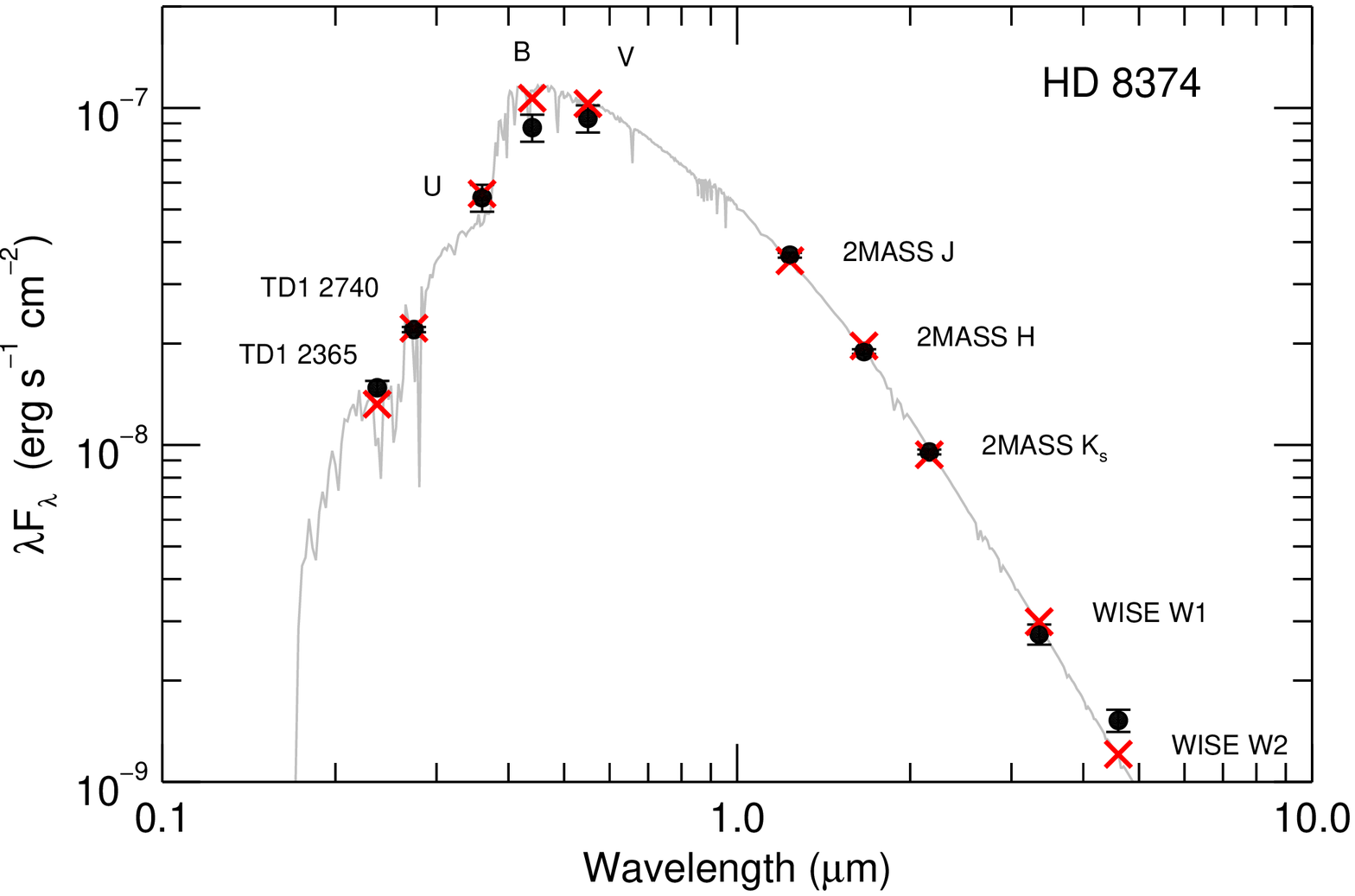} \hfill
\includegraphics[width=0.49\textwidth]{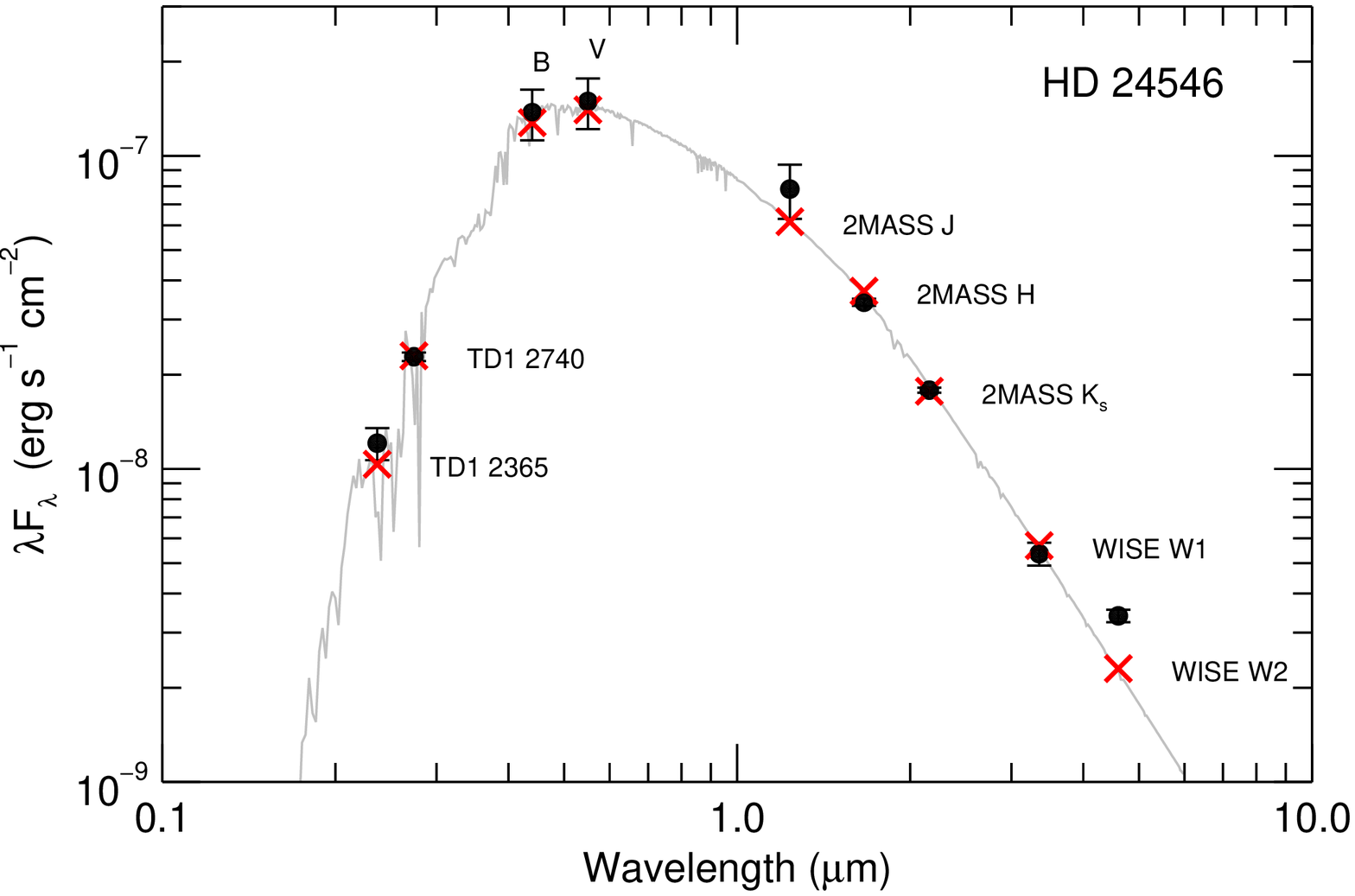} 
\caption{SED of HD~8374 (left) and HD~24546 (right). The observed fluxes are shown in black, and the best-fit binary model fluxes are shown as the red crosses. The full binary model is shown in gray. \label{sed}}
\end{figure*}

\section{Stellar Parameters}\label{section:param}

\subsection{Masses and Distance}
From the visual and spectroscopic orbital solution, we calculated the masses and distances of HD~8374 and HD~24546 using the nominal solar values from \citet{prsa16}. All of the fundamental stellar parameters for HD~8374 and HD~24546 are listed in Table~\ref{atmospar}.
We found 
HD~8374 to have masses of 
$M_1 = 1.636\pm0.050 M_{\odot}$ and 
$M_2 = 1.587\pm0.049 M_\odot$, and a distance of 
$d = 61.7\pm0.7$ pc.  
The distance from our orbital parallax is consistent with the \textit{Gaia} DR2 \citep{gaia1} distance of $62.5\pm0.6$ pc \citep{gaia2} from its trigonometric parallax.  
For HD~24546, we found masses of 
$M_1 = 1.434\pm0.014 M_{\odot}$ and 
$M_2 = 1.409\pm0.014 M_\odot$, and a distance of 
$d = 38.7\pm0.2$ pc.  
This is also consistent with the \textit{Gaia} DR2 distance of $38.4\pm0.2$ pc for HD~24546 \citep{gaia2}.

\begin{deluxetable}{lcc}
\tablewidth{0pt}
\tabletypesize{\footnotesize}
\tablecaption{Stellar Parameters  \vspace{6pt} \label{atmospar}}
\tablehead{ \colhead{Parameter} &\colhead{HD~8374} & \colhead{HD~24546} }
\startdata			
$M_1$ ($M_\odot$)	& $ 1.636\pm0.050 $		&  $1.434\pm0.014$ \\ 
$M_2$ ($M_\odot$)	& $ 1.587\pm0.049$     	&  $1.409\pm0.014$ \\ 
$R_1$ ($R_\odot$)	& $ 1.84\pm0.05 $     	&  $1.67\pm0.06$     \\ 
$R_2$ ($R_\odot$)	& $ 1.66\pm0.12 $     	&  $1.60\pm0.10$     \\ 
$T_{\rm eff \ 1}$ (K)	& $ 7280\pm110 $   		&  $6790\pm120$   \\ 
$T_{\rm eff \ 2}$ (K)	& $ 7280\pm120 $   		&  $6770\pm90$   \\ 
$\log g_1$ (cgs)	& $ 4.16\pm0.02$    		&  $4.15\pm0.02$ \\ 
$\log g_2$ (cgs)	& $ 4.22\pm0.03$    		&  $4.18\pm0.03$ \\ 
$V_1 \sin i$ (\kms)	& $ 15.9\pm1.3 $  		& $14.1\pm0.9$    \\ 
$V_2 \sin i$ (\kms)	& $ 15.2\pm1.4 $ 		& $10.6\pm0.7$    \\ 
Distance (pc)		& $ 61.7\pm0.7 $    	 	& $38.7\pm0.2$    \\
$E(B-V)$ (mag)		& $ 0.04\pm0.01 $    		& $0.07\pm0.02$  \\ 
\enddata                   
\end{deluxetable}

\break

\subsection{Effective Temperatures and Rotational Velocities}\label{tempfit}
To determine the atmospheric parameters of these stars, we first reconstructed the individual spectrum of each component using a Doppler tomography algorithm \citep{tomography} with BLUERED template spectra as inputs.  We then compared the reconstructed spectra to model spectra of various effective temperatures (\teff) and projected rotational velocities ($V \sin i$).  For each combination of \teff\ and $V \sin i$, we calculated the CCF of the model and reconstructed spectra at several echelle orders featuring strong metal absorption lines. We added the CCFs from all orders together to form a grid of CCFs as a function of \teff\ and $V \sin i$, then interpolated within the grid to find the CCF maximum position and the corresponding best-fit \teff\ and $V \sin i$ for each component.
For HD~8374, we found 
$T_{\rm eff \ 1} = 7280\pm110$ K, 
$T_{\rm eff \ 2} = 7280\pm120$ K,  
$V_1 \sin i = 15.9\pm1.3$ \kms, and 
$V_2 \sin i = 15.2\pm1.4$ \kms.
Both components are rotating faster than their pseudo-synchronous velocities of 9.6 \kms\ and 8.7 \kms.
For HD~24546, we found 
$T_{\rm eff \ 1} = 6790\pm120$ K, 
$T_{\rm eff \ 2} = 6770\pm90$ K,  
$V_1 \sin i = 14.1\pm0.9$ \kms, and 
$V_2 \sin i = 10.6\pm0.7$ \kms.
Both components are rotating close to their pseudo-synchronous velocities \citep{hilditch} of 12.9 \kms and 12.4 \kms.
Figure~\ref{recspec} shows the reconstructed spectra of each component and the model spectra created using these best-fit atmospheric parameters. Extended figure sets are included in the Appendix.

\begin{figure*}
\centering
\includegraphics[height=8cm]{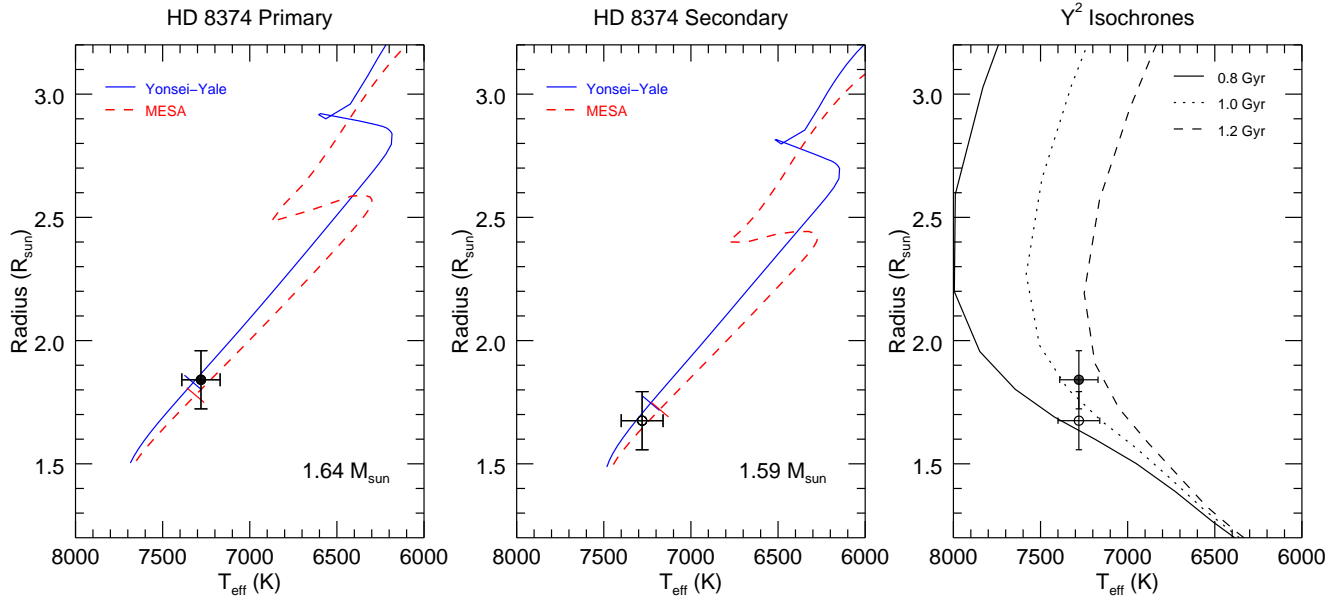} 
\caption{Evolutionary tracks (left, center) and isochrones (right) for HD~8374. The observed stellar parameters are shown as the filled points for the primary star and open points for the secondary star. The Yonsei-Yale models are shown as solid blue lines and the MESA models are shown as dashed red lines. The orthogonal tick marks represent the position of the mean system age on each track. The Yonsei-Yale $Y^2$ isochrones for 0.8, 1.0, and 1.2 Gyr are shown in the right panel. \label{evo8374}}
\end{figure*}

\break

\subsection{Radii and Surface Gravities}\label{sedfit}	
We gathered photometry from the literature to create spectral energy distributions (SEDs) of HD~8374 and HD~24546, including ultraviolet fluxes from TD-1 \citep{TD1}, optical fluxes from \citet{soubiran16}, and infrared fluxes from 2MASS \citep{2mass} and WISE \citep{wise}. We then created a binary SED model using 
$$f_\lambda = \frac{1}{d^2}  \Big(    R_1^2\ F_{\lambda 1} + R_2^2\ F_{\lambda 2} \Big) \times 10^{-0.4 A_\lambda}$$ 
where $F_{\lambda 1}$ and $F_{\lambda 2}$ are surface flux models of each component  \citep{sedmodel}, 
$R_1$ and $R_2$ are the stellar radii, $d$ is the distance, 
and $A_\lambda$ is the extinction in magnitudes. We used the reddening curves ($R_\lambda$) of \citet{redcurve} to calculate the extinction at each wavelength for a given color excess, where $A_\lambda = R_\lambda \times E(B-V).$
We also calculated the radius ratio ($R_2/R_1$) of each system from the model surface flux ratio and the observed flux ratios near H$\alpha$ (from the spectroscopic flux ratio) and in $K'$-band (from the interferometric flux ratio). The weighted-average radius ratio is then $R_2/R_1=0.91\pm0.06$ for HD~8374 and $R_2/R_1=0.96\pm0.05$ for HD~24546. 

We substituted this ratio into the above equation, then fit the binary model SED to the observed fluxes in order to determine $R_1$ and $A_\lambda$ (reported in terms of $E(B-V)$ in Table~\ref{atmospar}). We calculated $R_2$ from the radius ratios and calculated the surface gravities ($\log g$) from the masses and radii. 
For HD~8374, we found 
$R_1 = 1.84\pm0.05 R_\odot$,  
$R_2 = 1.66\pm0.12 R_\odot$, 
$\log g_1 = 4.16\pm0.02$ and 
$\log g_2 = 4.22\pm0.03$.  
For HD~24546, we found 
$R_1 = 1.67\pm0.06 R_\odot$, 
$R_2 = 1.60\pm0.10 R_\odot$, 
$\log g_1 = 4.15\pm0.02$ and 
$\log g_2 = 4.18\pm0.03$. 
The observed fluxes and the best-fit binary SED model are shown in Figure~\ref{sed}.

\begin{figure*}
\centering
\includegraphics[height=8cm]{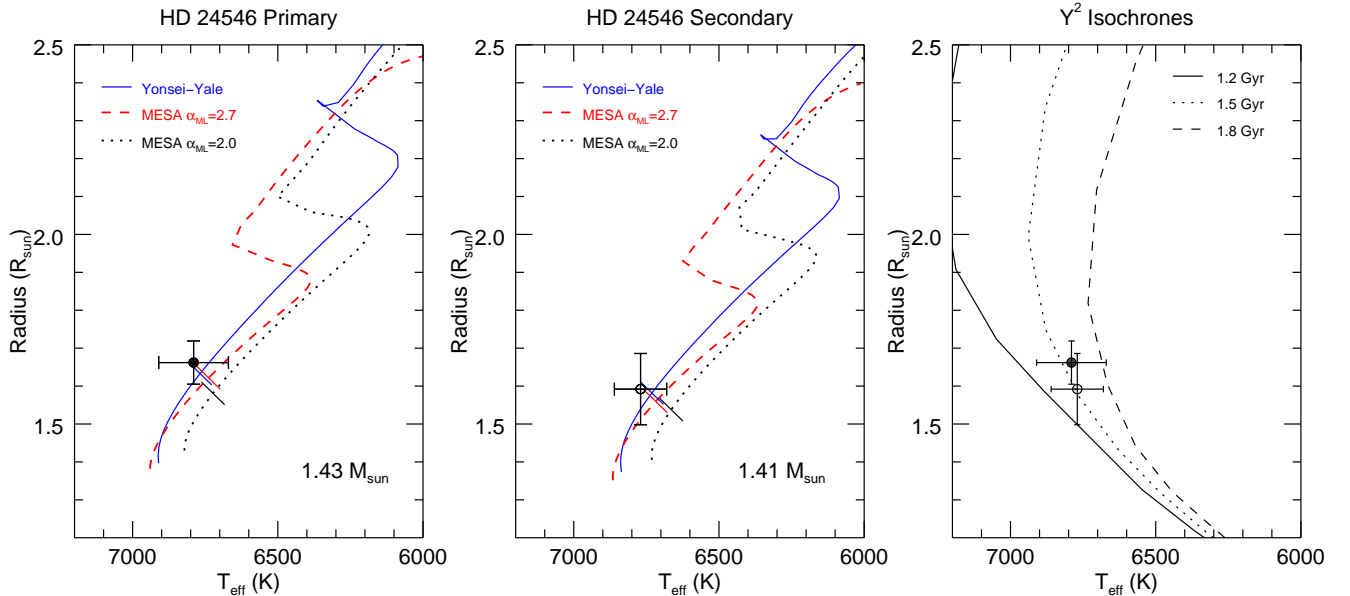}
\caption{Evolutionary tracks (left, center) and isochrones (right) for HD~24546. The observed stellar parameters are shown as the filled, black points for the primary star and open points for the secondary star. The Yonsei-Yale models are shown as solid blue lines. The MESA models for $\alpha_{\rm ML}=2.7$ are shown as dashed red lines, while the MESA  models for $\alpha_{\rm ML}=2.0$ are shown as dotted gray lines. The orthogonal tick marks represent the position of the mean system age on each track. The Yonsei-Yale isochrones for 1.2, 1.5, and 1.8~Gyr are shown in the right panel. \label{evo24546}}
\end{figure*}

\subsection{Comparison with Evolutionary Models} 
To estimate the ages of HD~8374 and HD~24546, we compared the observed stellar parameters to the predictions of the evolutionary models from Yonsei-Yale $Y^2$ \citep{y2} and MESA \citep{mesa1, mesa2, mesa3, mesa4, mesa5}.   The Yonsei-Yale $Y^2$ models were made for each component with their interpolation program\footnote{\href{http://www.astro.yale.edu/demarque/yystar.html}{http://www.astro.yale.edu/demarque/yystar.html}}. These models adopt a mixing length parameter of $\alpha_{\rm ML}$=1.74, which corresponds to the mixing length divided by the local pressure scale height.  The $Y^2$ models also use a step function prescription for the convective core overshooting, where the overshooting parameter ($\alpha_{\rm ov}$) increases from 0.0--0.2 based on the star's mass. 
 The MESA models\footnote{\href{http://www.mesa.sourceforge.net/index.html}{http://www.mesa.sourceforge.net/index.html}} were created at the observed masses with MESA release 10108. The default mixing length parameter is $\alpha_{\rm ML}$=2.0 for these models.  We also chose overshooting parameters of $f_{\rm ov}=0.01 $ for both components of HD~8374 and $f_{\rm ov}=0.005$ for both components of HD~24546 based on the empirical calibration of \citet{ct18}. Both sets of  models are also non-rotating and use solar metallicity.

Finally, we estimated the age of each binary system by averaging the individual component ages (where the model most closely matches the observed parameters), and determined the uncertainty from the range of ages that match the parameters of both components.   For HD~8374, we found the system age to be $1.0\pm0.1$ Gyr in the $Y^2$ models and $0.8\pm0.2$ Gyr in the MESA models. The evolutionary tracks and isochrones for HD~8374 are shown in Figure~\ref{evo8374}. The models successfully intersect the observed properties of both components at a single age.

\begin{figure*}
\centering
\includegraphics[height=8cm, trim = 0 0 350 0, clip]{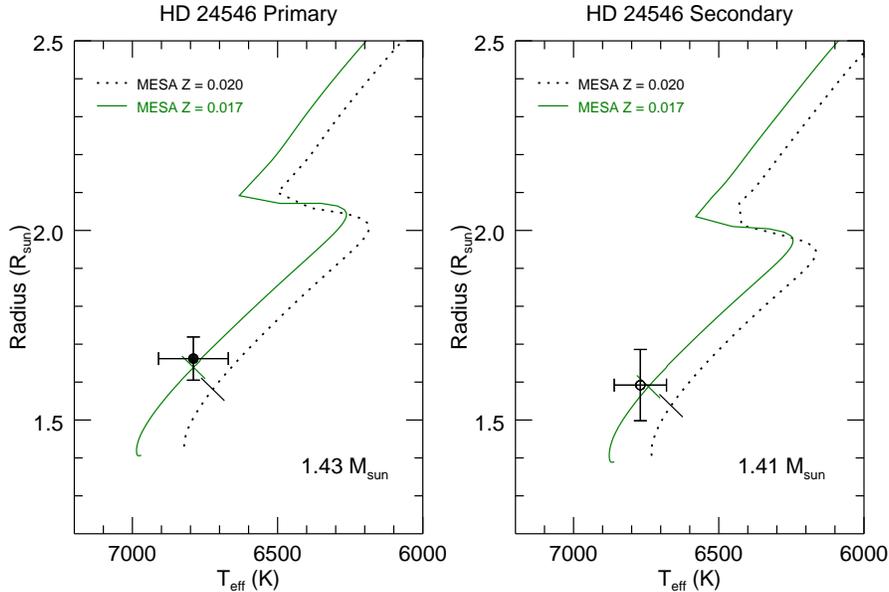} 
\caption{MESA evolutionary tracks for the primary component (left) and the secondary component (right) for two different metallicities. The $Z=0.017$ models are shown in solid green and the solar $Z=0.020$ models are shown as black dotted lines. The orthogonal tick marks represent the position of the mean system age on each track.   \label{evo24546b}}
\end{figure*}

For HD~24546, the $Y^2$ models successfully intersect the observations at an age of $1.5\pm0.3$~Gyr, as shown in Figure~\ref{evo24546}. However, the MESA models could not reproduce the observed values at the same age.  To solve this problem, we first tested different mixing length parameters from $\alpha_{\rm ML}=1.8 - 2.7$, the range tested by \citet{ct18}, to find which tracks intersect the observations with the least difference in age between the components. We found the best value to be $\alpha_{\rm ML}=2.7$ for both components and a corresponding system age of $1.4\pm0.2$~Gyr.  MESA evolutionary tracks for the default value of $\alpha_{ML} = 2.0$ and the best-fit value of $\alpha_{ML} =2.7$ are shown in Figure~\ref{evo24546}.   

Next, we kept $\alpha_{ML}$ fixed to 2.0 and tested different metallicities from $Z = 0.015 - 0.020$, where MESA adopts a solar metallicity of $Z_\odot=0.020$ from \citet{grevesse98}. For each metallicity, we refit for the component effective temperatures using the method described in Section~\ref{tempfit} and model spectra interpolated to the given metallicity.  Because the BLUERED spectra use $Z_\odot=0.019$ from \citet{anders89}, we used the same $\log(Z/Z_\odot)$ for both models and found that decreasing the ratio by $0.02$~dex resulted in a decreased effective temperature by about 50~K, which is within the temperature uncertainties.  We found that a slightly sub-solar metallicity of $Z=0.017$ could successfully fit the parameters of both components at an age of $1.3\pm0.2$~Gyr as shown in Figure~\ref{evo24546b}.

\section{Discussion}\label{section:discussion}

By measuring the visual and spectroscopic orbits of HD~8374 and HD~24546, we determined the masses of each component to within 3\% and the radii to within 7\%.  To better test the stellar evolution models, the next step would be to reduce the uncertainties in stellar radius. Typically, this is done with light curve modeling for short period, eclipsing systems. However, HD~8374 and HD~24546 have periods of 30 or more days, orbital inclinations far from 90 degrees, and radii less than twice that of the Sun, so neither HD~8374 nor HD~24546 is expected to show eclipses.

Therefore, better radius estimates can only be measured by resolving the component stars with long baseline interferometry at visible wavelengths. The PAVO \citep{pavo} and VEGA  \citep{vega} beam combiners at CHARA currently have angular resolutions of 0.25~mas  and 0.20~mas, respectively, and NPOI \citep{npoi} is expected to have an angular resolution of 0.15 mas when the 432~m baseline is installed. The components of HD~8374 have estimated angular diameters of 0.29 and 0.26 mas, while the components of HD~24546 have estimated angular diameters of 0.40 and 0.38 mas, so both arrays would be able to resolve the individual stars within these systems.

We also confirm that both components of HD~8374 show much weaker \ion{Ca}{2} lines than found in the model spectra of an early F-type star, as seen in Figure~\ref{8374figset}. This is consistent with other Am stars, which are defined by an apparent surface under-abundance of calcium \citep{conti70}. The metal and hydrogen line depths of HD~8374 appear to match the models, so these abundances are likely close to solar.

Finally, we report on whether or not HD~24546 is a member of the Hyades cluster. The cluster distance is $48.3\pm2.0$ pc at the center with an estimated radius of 10 pc \citep{perryman98}. We found the distance of HD~24546 to be $38.6\pm0.4$ pc, consistent with the inner edge of the Hyades cluster. However, our age for HD~24546 from the $Y^2$ evolutionary tracks is $1.5\pm0.3$~Gyr, compared to ages for the Hyades cluster of $625\pm50$~Myr using non-rotating models \citep{perryman98} or $750\pm100$~Myr using rotating models \citep{brandt15}. The BANYAN code of \citet{banyan} also reports 0\% probability of cluster membership using positions, proper motions, radial velocities, and parallaxes. Thus, we conclude that HD~24546 is not a member of the Hyades cluster.

\acknowledgments

{\footnotesize{
The authors would like to thank the staff at APO and CHARA for their invaluable support during observations, as well as Joel Eaton for collecting some of the TSU 2~m AST spectra. Institutional support has been provided from the GSU College of Arts and Sciences and the GSU Office of the Vice President for Research and Economic Development. Astronomy at Tennessee State University (TSU) is supported by the state of Tennessee through its Centers of Excellence Program. JDM acknowledges funding from NASA NNX09AB87G. 
This work is based in part upon observations obtained with the Apache Point Observatory 3.5-meter telescope, owned and operated by the Astrophysical Research Consortium;  the Georgia State University Center for High Angular Resolution Astronomy Array at Mount Wilson Observatory, supported by the National Science Foundation under Grants No. AST-1636624 and AST-1715788; the Palomar Testbed Interferometer, operated by the NASA Exoplanet Science Institute and the PTI collaboration and developed by the Jet Propulsion Laboratory, California Institute of Technology with funding provided from the National Aeronautics and Space Administration; and the High-Resolution Imaging instrument `Alopeke at Gemini-North (GN-2018B-FT-102), funded by the NASA Exoplanet Exploration Program and built at the NASA Ames Research Center by Steve B. Howell, Nic Scott, Elliott P. Horch, and Emmett Quigley.  
This work has also made use of the Jean-Marie Mariotti Center SearchCal service, the CDS Astronomical Databases SIMBAD and VIZIER, the Wide-field Infrared Survey Explorer, and the Two Micron All Sky Survey. 
}}

\facilities{APO:3.5m, CHARA, Gemini:North, PO:PTI, TSU:AST} 

\software{ Grid Search for Binary Stars \citep{schaefer16}, MESA \citep{mesa1},  RVFIT \citep{rvfit},  SearchCal \citep{searchcal}, TODCOR \citep{todcor1}, Y$^2$ models \citep{y2} }

\appendix
This appendix includes figure sets showing the reconstructed and model spectra of HD~8374 and HD~24546. Each figure contains panels for three different echelle orders featuring strong absorption lines in the range 3930--6640\AA. The model spectra were created from BLUERED models corresponding to the atmospheric parameters listed in Table~\ref{atmospar} and solar metallicity.

\vspace{1cm}

\figsetstart
\figsetnum{1}
\figsettitle{Reconstructed spectra of HD 8374}

\figsetend

\begin{figure*}
\centering
\includegraphics[width=0.95\textwidth]{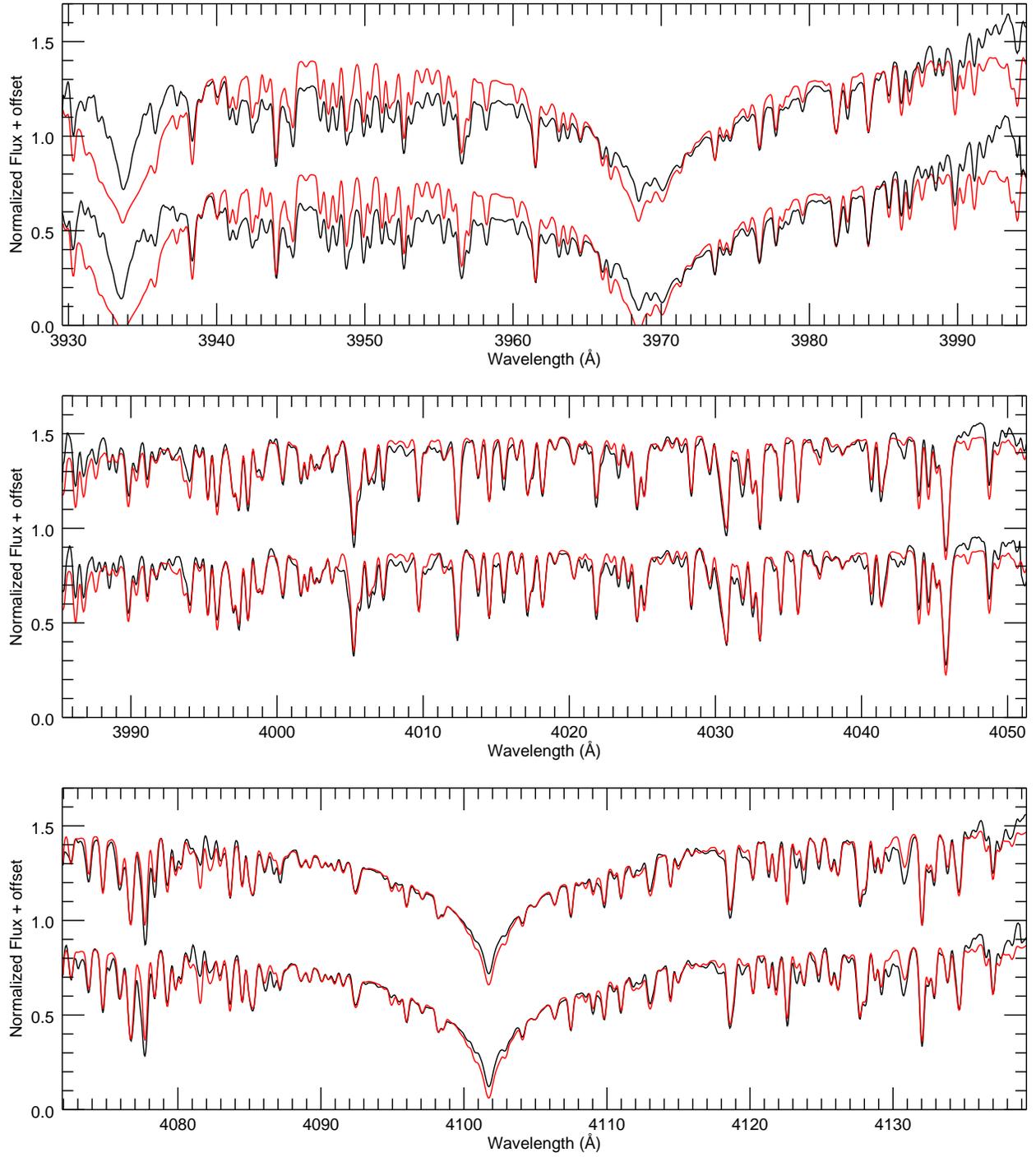}
\caption{Reconstructed spectra of HD 8374 are shown in black for the primary component (top) and secondary component (bottom). The best-fit model spectra are shown in red. The complete figure set (8 images) is available in the online journal. \label{8374figset}}
\end{figure*}

\figsetstart
\figsetnum{2}
\figsettitle{Reconstructed spectra of HD 24546}

\figsetend

\begin{figure*}
\centering
\includegraphics[width=0.95\textwidth]{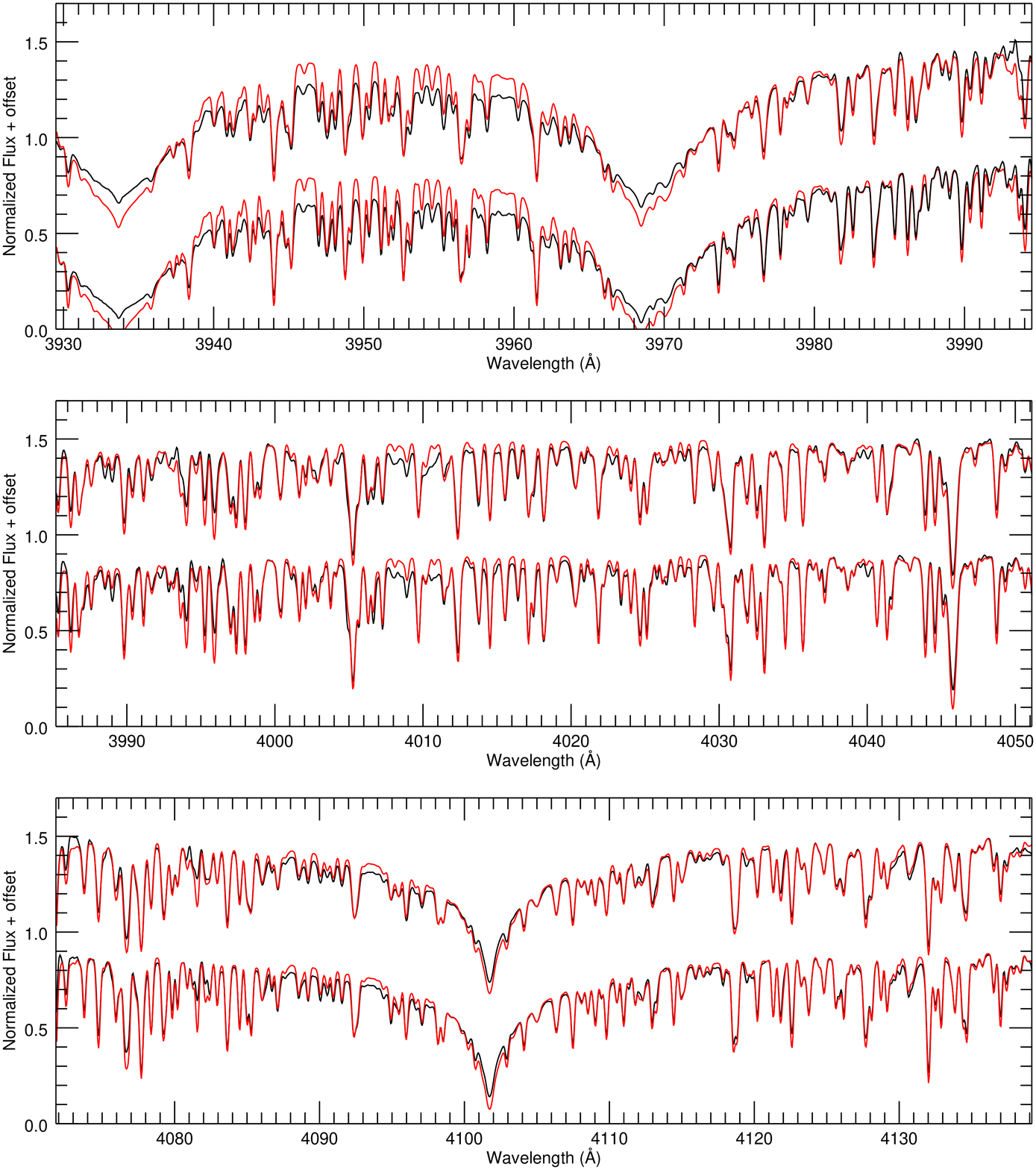}
\caption{Reconstructed spectra of HD 24546 are shown in black for the primary component (top) and secondary component (bottom). The best-fit model spectra are shown in red. The complete figure set (8 images) is available in the online journal.}
\end{figure*}

\clearpage



\begin{thebibliography}{}

\bibitem[Abt \& Levy(1976)]{abt76} 		
Abt, H.~A., \& Levy, S.~G.\ 1976, \apjs, 30, 273

\bibitem[Abt \& Morrell(1995)]{abt95} 	
Abt, H.~A., \& Morrell, N.~I.\ 1995, \apjs, 99, 135

\bibitem[Anders \& Grevesse(1989)]{anders89} 
Anders, E., \& Grevesse, N.\ 1989, \gca, 53, 197 

\bibitem[Armstrong et al.(1998)]{npoi} 
Armstrong, J.~T., Mozurkewich, D., Rickard, L.~J., et al.\ 1998, \apj, 496, 550

\bibitem[Bagnuolo et al.(1992)]{tomography} 
Bagnuolo, W.~G., Jr., Gies, D.~R., \& Wiggs, M.~S.\ 1992, \apj, 385, 708 

\bibitem[Bailer-Jones et al.(2018)]{gaia2} 
Bailer-Jones, C.~A.~L., Rybizki, J., Fouesneau, M., et al.\ 2018, \aj, 156, 58

\bibitem[Bertone et al.(2008)]{bluered} 
Bertone, E., Buzzoni, A., Ch{\'a}vez, M., \& Rodr{\'{\i}}guez-Merino, L.~H.\ 2008, \aap, 485, 823  

\bibitem[Boden et al.(1999)]{boden99} 
Boden, A.~F., Koresko, C.~D., van Belle, G.~T., et al.\ 1999, \apj, 515, 356

\bibitem[Boden et al.(2006)]{boden06} 
Boden, A.~F., Torres, G., \& Latham, D.~W.\ 2006, \apj, 644, 1193

\bibitem[Brandt \& Huang(2015)]{brandt15} 
Brandt, T.~D., \& Huang, C.~X.\ 2015, \apj, 807, 58

\bibitem[Castelli \& Kurucz(2004)]{sedmodel} 
Castelli, F., \& Kurucz, R.~L.\ 2004, arXiv:astro-ph/0405087, \href{http://adsabs.harvard.edu/abs/2004astro.ph..5087C}{ADS link}

\bibitem[Chelli et al.(2016)]{searchcal} 
Chelli, A., Duvert, G., Bourg{\`e}s, L., et al.\ 2016, \aap, 589, A112 

\bibitem[Claret \& Torres(2018)]{ct18}   
Claret, A., \& Torres, G.\ 2018, \apj, 859, 100 

\bibitem[Colavita et al.(1999)]{pti1} 	
Colavita, M.~M., Wallace, J.~K., Hines, B.~E., et al.\ 1999, \apj, 510, 505

\bibitem[Colavita(1999)]{pti2} 		
Colavita, M.~M.\ 1999, \pasp, 111, 111

\bibitem[Conti(1970)]{conti70} Conti, P.~S.\ 1970, \pasp, 82, 781

\bibitem[Demarque et al.(2004)]{y2} 
Demarque, P., Woo, J.-H., Kim, Y.-C., \& Yi, S.~K.\ 2004, \apjs, 155, 667 

\bibitem[Eaton \& Williamson(2004)]{eaton04} Eaton, J.~A., \& Williamson, M.~H.\ 2004, in Advanced Software, Control, and Communication Systems for Astronomy, Proc. SPIE, ed. H.~Lewis \& G. Raffi (Bellingham, WA: SPIE), 5496, 710

\bibitem[Eaton \& Williamson(2007)]{eaton07} 	
Eaton, J.~A., \& Williamson, M.~H.\ 2007, \pasp, 119, 886

\bibitem[Eggen(1971)]{eggen71} 
Eggen, O.~J.\ 1971, \pasp, 83, 741

\bibitem[Fekel et al.(2009)]{fekel09} 		
Fekel, F.~C., Tomkin, J., \& Williamson, M.~H.\ 2009, \aj, 137, 3900

\bibitem[Fekel et al.(2011)]{fekel11} 		
Fekel, F.~C., Tomkin, J., Williamson, M.~H., et al.\ 2011, \aj, 142, 69

\bibitem[Fekel et al.(2013)]{fekel13} 
Fekel, F.~C., Rajabi, S., Muterspaugh, M.~W., et al.\ 2013, \aj, 145, 111

\bibitem[Fitzpatrick(1999)]{redcurve} Fitzpatrick, E.~L.\ 1999, \pasp, 111, 63

\bibitem[Fletcher(1967)]{fletcher67} 
Fletcher, J.~M.\ 1967, \jrasc, 61, 56

\bibitem[Gagn{\'e} et al.(2018)]{banyan} 
Gagn{\'e}, J., Mamajek, E.~E., Malo, L., et al.\ 2018, \apj, 856, 23 

\bibitem[Gaia Collaboration et al.(2016)]{gaia1} 
Gaia Collaboration, Prusti, T., de Bruijne, J.~H.~J., et al.\ 2016, \aap, 595, A1 

\bibitem[Grevesse \& Sauval(1998)]{grevesse98} 
Grevesse, N., \& Sauval, A.~J.\ 1998, \ssr, 85, 161

\bibitem[Halbwachs(1981)]{halbwachs81} 
Halbwachs, J.~L.\ 1981, \aaps, 44, 47 

\bibitem[He{\l}miniak et al.(2012)]{heminiak12} 
He{\l}miniak, K.~G., Konacki, M., Muterspaugh, M.~W., et al.\ 2012, \mnras, 419, 1285

\bibitem[Hilditch(2001)]{hilditch} Hilditch, R.~W.\ 2001, An Introduction to Close Binary Stars (Cambridge, UK: Cambridge University Press), \href{https://ui.adsabs.harvard.edu/abs/2001icbs.book.....H/abstract}{ADS link}

\bibitem[Horch et al.(2017)]{horch17} 
Horch, E.~P., Casetti-Dinescu, D.~I., Camarata, M.~A., et al.\ 2017, \aj, 153, 212 

\bibitem[Howell et al.(2011)]{howell11} 
Howell, S.~B., Everett, M.~E., Sherry, W., Horch, E., \& Ciardi, D.~R.\ 2011, \aj, 142, 19 

\bibitem[Ireland et al.(2008)]{pavo} 
Ireland, M.~J., M{\'e}rand, A., ten Brummelaar, T.~A., et al.\ 2008, in Optical and Infrared Interferometry, \procspie, ed. M.~Schöller, W.~C.~Danchi, \& F.~Delplancke (Bellingham, WA: SPIE), 7013

\bibitem[Iglesias-Marzoa et al.(2015)]{rvfit} 
Iglesias-Marzoa, R., L{\'o}pez-Morales, M., \& Jes{\'u}s Ar{\'e}valo Morales, M.\ 2015, \pasp, 127, 567 

\bibitem[Kolbas et al.(2015)]{kolbas15}  
Kolbas, V., Pavlovski, K., Southworth, J., et al.\ 2015, \mnras, 451, 4150 

\bibitem[Konacki \& Lane(2004)]{konacki04} 
Konacki, M., \& Lane, B.~F.\ 2004, \apj, 610, 443

\bibitem[Kluska et al.(2018)]{kluska18}  
Kluska, J., Kraus, S., Davies, C.~L., et al.\ 2018, \apj, 855, 44 

\bibitem[L{\'e}pine \& Bongiorno(2007)]{lepine07} 	
L{\'e}pine, S., \& Bongiorno, B.\ 2007, \aj, 133, 889

\bibitem[Lester et al.(2019a)]{lester19a} 
Lester, K.~V., Gies, D.~R., Schaefer, G.~H., et al.\ 2019a, \aj, 157, 140

\bibitem[Lester et al.(2019b)]{lester19b} 
Lester, K.~V., Gies, D.~R., Schaefer, G.~H., et al.\ 2019b, \aj, 158, 218

\bibitem[Monnier et al.(2011)]{monnier11}   
Monnier, J.~D., Zhao, M., Pedretti, E., et al.\ 2011, \apjl, 742, L1 

\bibitem[Montes et al.(2018)]{montes18}     
Montes, D., Gonz{\'a}lez-Peinado, R., Tabernero, H.~M., et al.\ 2018, \mnras, 479, 1332

\bibitem[Mourard et al.(2009)]{vega} 
Mourard, D., Clausse, J.~M., Marcotto, A., et al.\ 2009, \aap, 508, 1073

\bibitem[Paxton et al.(2011)]{mesa1} 
Paxton, B., Bildsten, L., Dotter, A., et al.\ 2011, \apjs, 192, 3 

\bibitem[Paxton et al.(2013)]{mesa2} 
Paxton, B., Cantiello, M., Arras, P., et al.\ 2013, \apjs, 208, 4 

\bibitem[Paxton et al.(2015)]{mesa3} 
Paxton, B., Marchant, P., Schwab, J., et al.\ 2015, \apjs, 220, 15 

\bibitem[Paxton et al.(2018)]{mesa4} 
Paxton, B., Schwab, J., Bauer, E.~B., et al.\ 2018, \apjs, 234, 34 

\bibitem[Paxton et al.(2019)]{mesa5} 
Paxton, B., Smolec, R., Schwab, J., et al.\ 2019, \apjs, 243, 10

\bibitem[Perryman et al.(1998)]{perryman98} 
Perryman, M.~A.~C., Brown, A.~G.~A., Lebreton, Y., et al.\ 1998, \aap, 331, 81

\bibitem[Pr{\v{s}}a et al.(2016)]{prsa16} 	
Pr{\v{s}}a, A., Harmanec, P., Torres, G., et al.\ 2016, \aj, 152, 41

\bibitem[Sandberg Lacy \& Fekel(2011)]{sandberg11} 
Sandberg Lacy, C.~H., \& Fekel, F.~C.\ 2011, \aj, 142, 185

\bibitem[Scarfe(2010)]{scarfe10} 
Scarfe, C.~D.\ 2010, The Observatory, 130, 214

\bibitem[Schaefer et al.(2016)]{schaefer16} 
Schaefer, G.~H., Hummel, C.~A., Gies, D.~R., et al.\ 2016, \aj, 152, 213 

\bibitem[Scott et al.(2018)]{scott18} 
Scott, N.~J., Howell, S.~B., Horch, E.~P., et al.\ 2018, \pasp, 130, 054502

\bibitem[Skrutskie et al.(2006)]{2mass} 
Skrutskie, M.~F., Cutri, R.~M., Stiening, R., et al.\ 2006, \aj, 131, 1163

\bibitem[Soubiran et al.(2016)]{soubiran16}	
Soubiran, C., Le Campion, J.-F., Brouillet, N., et al.\ 2016, \aap, 591, A118

\bibitem[ten Brummelaar et al.(2005)]{chara} 
ten Brummelaar, T.~A., McAlister, H.~A., Ridgway, S.~T., et al.\ 2005, \apj, 628, 453 

\bibitem[ten Brummelaar et al.(2013)]{climb} 
ten Brummelaar, T.~A., Sturmann, J., Ridgway, S.~T., et al.\ 2013, 
Journal of Astronomical Instrumentation, 2, 1340004 

\bibitem[Thompson et al.(1978)]{TD1} 
Thompson, G.~I., Nandy, K., Jamar, C., et al.\ 1978. ``Catalogue of stellar ultraviolet fluxes: a compilation of absolute stellar fluxes measured by the Sky Survey Telescope (S2/68) aboard the ESRO satellite TD-1",  (London: The Science Research Council), \href{http://adsabs.harvard.edu/cgi-bin/bib_query?1978csuf.book.....T}{ADS link}

\bibitem[Tokovinin(1997)]{tokovinin97} 	
Tokovinin, A.~A.\ 1997, \aaps, 124, 75

\bibitem[van Belle et al.(2008)]{vanbelle08} 
van Belle, G.~T., van Belle, G., Creech-Eakman, M.~J., et al.\ 2008, \apjs, 176, 276

\bibitem[Wallerstein(1973)]{wallerstein73} 
Wallerstein, G.\ 1973, \pasp, 85, 115

\bibitem[Wang et al.(2003)]{arces} 
Wang, S.-i., Hildebrand, R.~H., Hobbs, L.~M., et al.\ 2003, in Instrument Design and Performance for Optical/Infrared Ground-based Telescopes, Proc. SPIE 4841, ed. M. Iye \& A. F. M. Moorwood (Bellingham, WA: SPIE), 1145

\bibitem[Wright et al.(2010)]{wise} 
 Wright, E.~L., Eisenhardt, P.~R.~M., Mainzer, A.~K., et al.\ 2010, \aj, 140, 1868 

\bibitem[Zucker \& Mazeh(1994)]{todcor1} 
Zucker, S., \& Mazeh, T.\ 1994, \apj, 420, 806 

\bibitem[Zucker et al.(2003)]{todcor2} 
Zucker, S., Mazeh, T., Santos, N.~C., Udry, S., \& Mayor, M.\ 2003, \aap, 404, 775 


\end{thebibliography}
\end{document}